\documentclass[11pt]{article}

\usepackage[preprint]{acl}
\usepackage{tikz}
\usepackage{times}
\usepackage{latexsym}
\usepackage{booktabs}
\usepackage{multicol}
\usepackage{multirow}
\usepackage{makecell}
\usepackage{listings}
\usepackage{xcolor}  
\usepackage{cuted}
\definecolor{dkgreen}{rgb}{0,0.6,0}
\definecolor{gray}{rgb}{0.5,0.5,0.5}
\definecolor{mauve}{rgb}{0.58,0,0.82}
\usepackage{amssymb}    
\usepackage{array}      
\usepackage{pifont} 
\newcommand{\cmark}{\ding{51}}%
\newcommand{\xmark}{\ding{55}}%
\usepackage[T1]{fontenc}

\usepackage[utf8]{inputenc}

\usepackage{microtype}

\usepackage{inconsolata}

\usepackage{graphicx}
\usepackage{tablefootnote}

\newcommand{\vpara}[1]{\vspace{0.5em}\noindent\textbf{#1}\quad}

\usepackage[utf8]{inputenc}
\usepackage{tcolorbox}
\tcbuselibrary{breakable, skins}
\newtcolorbox{promptbox}[1][]{
    enhanced,
    colback=gray!2,           
    colframe=black,          
    boxrule=1.5pt,           
    arc=4pt,                 
    outer arc=4pt,
    drop shadow={black!50},  
    title={#1},              
    fonttitle=\bfseries,     
    colbacktitle=black,      
    coltitle=white,          
    left=5pt, right=5pt, top=5pt, bottom=5pt, 
    breakable,               
    before skip=10pt, after skip=10pt,
}

\usepackage{caption}
\usepackage{subcaption} 
\title{PlotGen-Bench: Evaluating VLMs on Generating Visualization Code from Diverse Plots across Multiple Libraries}
\vspace{-10mm}
\author{
 \textbf{Yi Zhao\textsuperscript{1*}},
 \textbf{Zhen Yang\textsuperscript{1*}},
 \textbf{Shuaiqi Duan\textsuperscript{2}},
 \textbf{Wenmeng Yu\textsuperscript{2}},
 \textbf{Zhe Su\textsuperscript{2}},
 \\
 \textbf{Jibing Gong\textsuperscript{3$\dagger$}},
 \textbf{Jie Tang\textsuperscript{1$\dagger$}},
\\
 \textsuperscript{1}Tsinghua University,
 \textsuperscript{2}Zhipu AI,
 \textsuperscript{3}Yanshan University
}
\begin{document}
\maketitle
\mbox{} 
\vspace{-15mm}
\begin{abstract}
Recent advances in vision–language models (VLMs) have expanded their multimodal code generation capabilities, yet their ability to generate executable visualization code from plots, especially for complex 3D, animated, plot-to-plot transformations, or multi-library scenarios, remains underexplored. To address this gap, we introduce PlotGen-Bench, a comprehensive benchmark for evaluating plot-to-code generation under realistic and complex visualization scenarios. The benchmark spans 9 major categories, 30 subcategories, and 3 core tasks—plot replication, plot transformation, and multi-library generation, covering both 2D, 3D and animated plots across 5 widely used visualization libraries. Through systematic evaluation of state-of-the-art open- and closed-source VLMs, we find that open-source models still lag considerably behind in visual fidelity and semantic consistency, despite achieving comparable code executability. Moreover, all models exhibit substantial degradation on reasoning-intensive tasks such as chart type conversion and animation generation. PlotGen-Bench establishes a rigorous foundation for advancing research toward more capable and reliable VLMs for visualization authoring and code synthesis, with all data and code available at https://plotgen.github.io.
\end{abstract}
\let\thefootnote\relax\footnotetext{
$^*$ Equal contributions}

{\let\thefootnote\relax\footnotetext{
$^\dagger$ Corresponding authors}
}
\vspace{-4mm}
\section{Introduction}
Recent advances in vision-language models (VLMs) have significantly expanded their capabilities across text, vision, and other modalities, enabling applications such as multimodal reasoning~\cite{lu2022learn,achiam2023gpt,huang2025gemini}, graphical user interface (GUI) agents~\cite{hong2024cogagent,ye2025mobile}, and visual question answering~\cite{shao2023prompting,kim2025visual}. Among the many modalities that VLMs must handle, data visualizations stand out as a particularly important and underexplored domain. Visualizations are central to data analysis, scientific reporting, and decision making, serving both as a medium of communication and a tool for exploration. While prior work has demonstrated that VLMs can perform a range of chart understanding tasks including optical character recognition (OCR)~\cite{yang2025qwen3,chen2025logics}, text extraction~\cite{wu2024deepseek,wang2024cogvlm}, and reasoning over visualizations~\cite{yang2025qwen3,vteam2025glm45vglm41vthinkingversatilemultimodal,hong2024cogvlm2}, the equally critical ability to generate executable visualization code from plots remains largely underexplored.

\begin{figure}[t!]
    \vspace{-7mm}
    \centering
    \includegraphics[width=\linewidth]{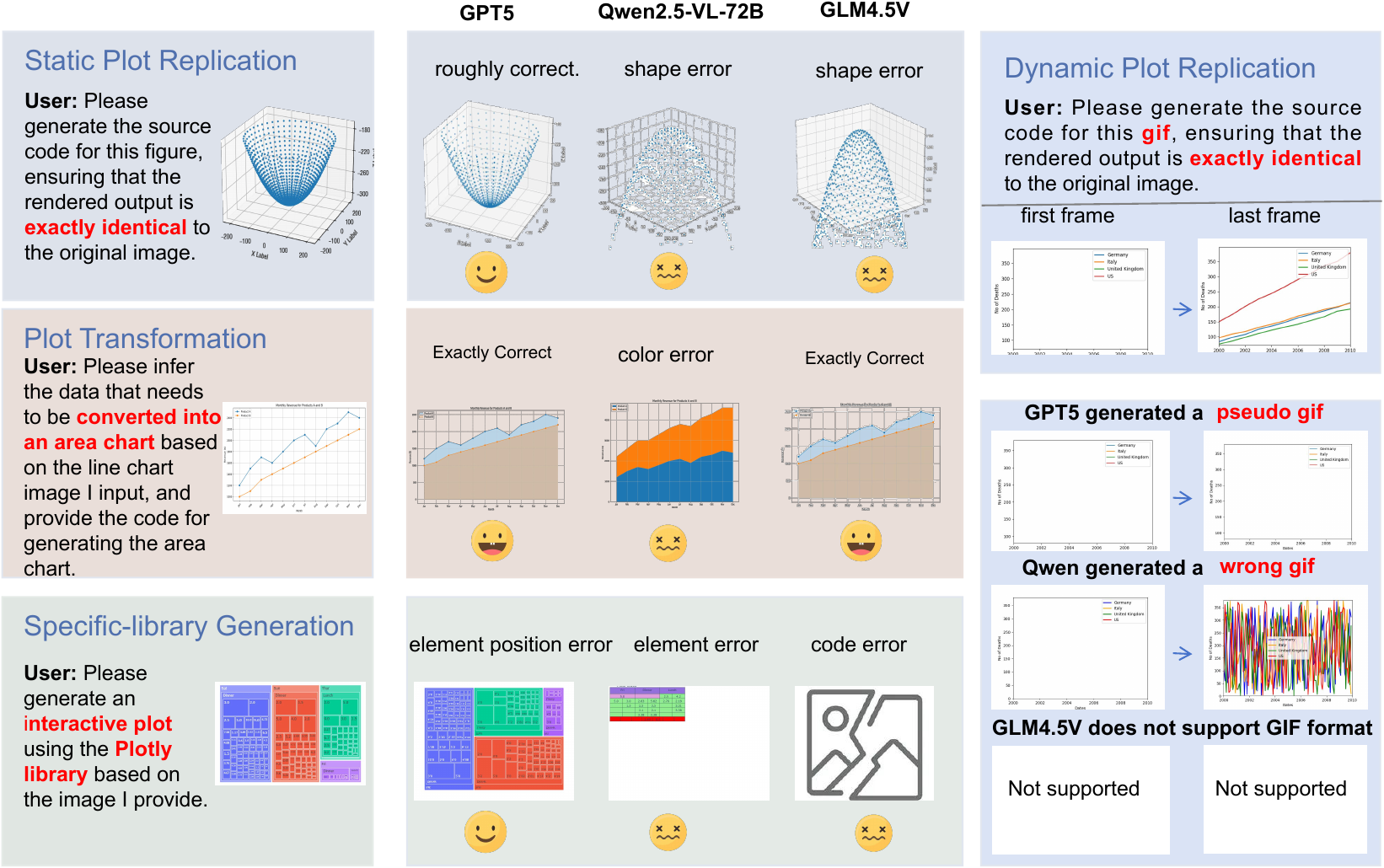}
    \caption{Comparison between regular figures on different libraries. and 3D, animation, interactive transformation on different models.}
    \vspace{-6mm}
    \label{fig:generate_result}
\end{figure}

In practice, researchers and practitioners often need not only to interpret a plot but also to reproduce, edit, or reuse it through executable code, which is an essential step toward scientific reproducibility and interactive visualization authoring. As shown in Figure \ref{fig:generate_result}, current VLMs struggle with diverse and complex visualization scenarios, including 3D plots, animations, and cross-library conversions. This underscores the lack of a benchmark that systematically evaluates plot-to-code generation beyond simple 2D reproduction. 

Existing benchmarks such as ChartMimic~\cite{yang2024chartmimic}, Plot2code~\cite{wu2024plot2code} provide an initial step toward plot-to-code evaluation, but they remain limited in scope. Specifically, current benchmarks (i) focus mainly on 2D static plots, (ii) are restricted to a single visualization library (e.g., Matplotlib), and (iii) evaluate only direct reproduction without considering richer tasks such as plot transformations. As a result, they fall short of capturing the diverse and practical scenarios in which users interact with plots, such as converting one visualization type into another (e.g., bar $\rightarrow$ line), or reproducing the plot across different libraries. Moreover, modern visualization practice frequently involves 3D and animated plots, which existing benchmarks do not capture and which present unique evaluation challenges. More detailed comparative information can be found in Table \ref{tab:dataset_comparison}.

To overcome these limitations, we introduce PlotGen-Bench, a comprehensive benchmark designed to systematically evaluate VLMs on diverse plot-to-code generation tasks. By covering 28 plot types, 3D and animated visualizations, multiple Python libraries (Matplotlib, Plotly, Seaborn, Networkx, Plotnine), and introducing tasks such as plot transformation, PlotGen-Bench goes beyond existing efforts to provide a richer and more realistic evaluation setting. In addition, our benchmark contributes a novel evaluation pipeline that combines automatic checks with VLM-as-a-judge assessments, enabling robust measurement across visual fidelity, executability, and semantic correctness.

In summary, our contributions are threefold:
\begin{itemize}
    \item \textbf{Comprehensive and Diverse Benchmark:} We present PlotGen-Bench, a large-scale benchmark covering 28 plot types, 3D and animated visualizations, and tasks such as plot transformation, reflecting the complex and practical needs of real-world visualization authoring.

    \item \textbf{Multiple Libraries Support:} Unlike prior work restricted to a single library, PlotGen-Bench supports multiple Python visualization libraries, including Matplotlib, Plotly, Seaborn, NetworkX, and Plotnine—thus exposing challenges of cross-framework generalization.

    \item \textbf{Robust Evaluation Pipeline:} We design a novel evaluation framework that combines automatic checks (executability, library compliance, visual similarity) with VLM-as-a-judge assessments, enabling both low-level and high-level evaluation of plot-to-code generation.
\end{itemize}

\begin{table*}[ht]
\centering
\small
\resizebox{\textwidth}{!}{
\begin{tabular}{l c c c c c c c c  c c}
\toprule
\textbf{Dataset} &
\multicolumn{2}{c}{\textbf{Input Type}} & 
\multicolumn{5}{c}{\textbf{Libraries}} & 
\multicolumn{3}{c}{\textbf{Task}} \\
\cmidrule(lr){2-3} \cmidrule(lr){4-8} \cmidrule(lr){9-11}
 & static png & gif & Matplotlib & Seaborn & Plotly & Plotnine & NetworkX & Plot Replication & Multi-Library Generation & Plot Transformation  \\
\midrule
Plot2Code \cite{wu2024plot2code} & \cmark & \xmark & \cmark & \xmark & \xmark & \xmark & \xmark & \cmark & \xmark & \xmark\\
ChartMimic \cite{yang2024chartmimic} & \cmark & \xmark & \cmark & \xmark & \xmark & \xmark & \xmark & \cmark & \xmark & \xmark\\
ChartEdit \cite{zhao2025chartedit}& \cmark & \xmark & \cmark & \cmark & \xmark & \xmark & \xmark & \cmark & \cmark & \xmark\\
ChartX \cite{xia2025chartx} & \cmark & \xmark & \cmark & \xmark & \xmark & \xmark & \xmark & \cmark & \xmark & \xmark\\
ChartM$^3$ \cite{yang2025chartm3benchmarkingchartediting} & \cmark & \xmark & \cmark & \xmark & \xmark & \xmark & \xmark & \xmark & \xmark & \cmark\\
\midrule
PlotGen-Bench & \cmark & \cmark & \cmark & \cmark & \cmark & \cmark & \cmark & \cmark & \cmark & \cmark\\
\bottomrule
\end{tabular}
}
\caption{Comparison of benchmark across input types, libraries, and task. (\cmark: Supported, \xmark: Not Supported)}
\vspace{-5mm}
\label{tab:dataset_comparison}
\end{table*}

\vspace{-4mm}
\section{Related Work}
\subsection{Vision-language Models For Plot Code Generation}
Recent advances in vision-language models have shifted plot understanding from pipeline designs \cite{methani2020plotqa} toward end-to-end generation of executable code \cite{openai2024gpt4technicalreport,bai2025qwen2,wang2025internvl3,vteam2025glm45vglm41vthinkingversatilemultimodal,young2024yi}, which serves as a near-lossless representation of visual structure. In charts, natural-language descriptions often omit dense layout and styling cues; by contrast, producing plotting scripts (e.g., Matplotlib/Plotly) preserves axes, marks, legends, fonts, and color maps in a machine-verifiable form. ChartCoder \cite{zhao2025chartcoder} pushes this view further by using a code-centric backbone and introducing Chart2Code-160k plus “Snippet-of-Thought” supervision, yielding notable gains in executability and structural restoration on chart-to-code tasks. ChartVLM \cite{xia2024chartx} establishes a chart-specialized VLM for complex reasoning, highlighting the remaining gap between high-level reasoning and faithful code synthesis. ChartMaster \cite{tan2025chartmaster} briefly extends this line by using real-world chart prompts and a visual-similarity-guided RL objective, yielding competitive results among 7B models. This body of work motivates moving beyond text-only QA toward evaluation centered on runnable code and reconstruction fidelity.

\subsection{Vision-language Models Code Generation Benchmark}
Recent benchmarks \cite{yang2024chartmimic,zhao2025chartedit,li2024mmcode,wu2024plot2code,wang2024charxiv} examine whether models can translate visual inputs into executable Python code, evaluating both the produced scripts and the rendered figures. For example, ChartMimic \cite{yang2024chartmimic} focuses on chart and instruction to code generation with a curated collection that spans many chart families and introduces multi level automatic scoring for code and image outputs. MMCode \cite{li2024mmcode} broadens the setting to visually conditioned programming tasks, supplying thousands of problems with images and running the generated Python solutions against hidden tests to assess executable correctness in realistic conditions. 
Complementing these resources, Plot2Code \cite{wu2024plot2code} centers its evaluation on three indicators: pass rate for compilability and execution, text match ratio for code level correspondence, and a GPT-based judgment that provides a holistic rating of the rendered output, which together form a comprehensive measure of chart-to-code performance.  ChartEdit \cite{zhao2025chartedit} shifts the target from generation to editing, pairing real charts with human-verified edit instructions and reference Python code to measure execution rate, code quality, and chart level fidelity. 
However, these benchmarks provide limited coverage of 3D charts and animated visualizations, despite their prevalence in practice. In addition, they rarely assess how models perform under different Python plotting frameworks such as Matplotlib, Seaborn, Plotly, Plotnine, and NetworkX.

\section{PlotGen-Bench}
In this section, we introduce \textbf{PlotGen-Bench}, a new benchmark designed to systematically evaluate visualization generation across diverse plot types, libraries, and tasks. 

\subsection{Scope and Task Definition}
The primary goal of PlotGen-Bench is to provide a comprehensive and systematic evaluation of visualization generation across a broad range of realistic plot types, visualization libraries, and generation tasks. Different from previous benchmarks that primarily focus on simple 2D static plots rendered in a single library, PlotGen-Bench is designed to capture the diversity and complexity of real-world visualization practice.

\begin{figure}
    \centering
    \includegraphics[width=1\linewidth]{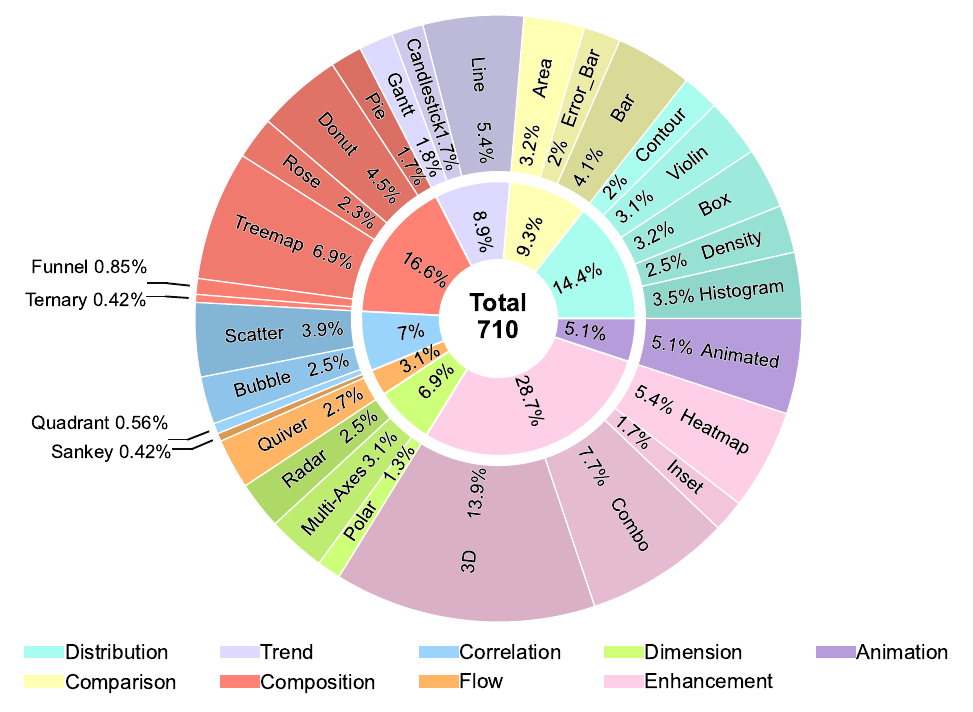}
    \caption{Plot subcategories  distribution in PlotGen-Bench}
    \label{fig:plot_distribution}
    \vspace{-6mm}
\end{figure}

\vpara{Scope of plot types.} PlotGen-Bench covers 30 distinct plot types, spanning conventional 2D plots (e.g., bar, line, scatter, pie), advanced visualization families (e.g., heatmaps, inset, polar, rose.), as well as 3D plots, interactive plots, and animated plots. To provide a systematic taxonomy of visualization forms, the 30 plot types in PlotGen-Bench are organized into 9 high-level categories grounded in their analytical intent and graphical representation patterns.
A detailed quantitative breakdown of these categories is illustrated in Figure~\ref{fig:plot_distribution}.

\vpara{Task definition.} 
PlotGen-Bench is organized around 3 key tasks that together represent practical scenarios of plot-to-code generation:
\begin{itemize}
    \item \textbf{Plot Replication: } Given an input plot image, VLMs are required to generate executable code that accurately reconstructs the visualization, preserving its layout, visual style, and data attributes. 
    \item \textbf{Plot Transformation: } Given an input plot image, VLMs are required to transform the visualization into another valid plot type (e.g., bar $\rightarrow$ line) while preserving the underlying data relationships, thereby assessing their ability for cross-type visual generation.
    \item \textbf{Multi-library Generation: } To evaluate generation across diverse visualization environments, PlotGen-Bench incorporates multiple Python visualization libraries, including \textit{Matplotlib}, \textit{Seaborn}, \textit{Plotly}, \textit{NetworkX}, and \textit{Plotnine}, encompassing both static and interactive rendering paradigms. 
\end{itemize}
By integrating these tasks, PlotGen-Bench extends beyond conventional plot-to-code generation to provide a unified benchmark covering the diverse challenges of real-world visualization generation.

\subsection{Data Construction}

\textbf{Data Sources and Collection.} 

To construct PlotGen-Bench, we curated diverse visualization samples from multiple high-quality sources to ensure realism and coverage. For common static and animated plots, we collected examples from open-source repositories and official tutorial of Matplotlib, Seaborn, and NetworkX. To include advanced forms such as interactive plots, we incorporated samples from Plotly and Plotnine galleries, as well as community platforms like Stack Overflow, Kaggle, and Zhihu. Additionally, we manually selected figures from arXiv papers to capture research-oriented visualizations with customized layouts and complex semantics. This multi-source process yielded a dataset spanning 30 plot types, 9 categories, and 5 libraries, providing a comprehensive foundation for evaluating visualization generation across real-world scenarios.

\vpara{Plot Quality Filtering.}
To ensure the visual and informational quality of the collected plots, we employ a two-stage filtering pipeline combining automated and human evaluation. First, GPT-4o \cite{openai2024gpt4technicalreport} automatically assesses each plot and its prompt for aesthetic quality, information density, and sensitive or non-compliant content, removing low-quality or redundant samples. Then, human annotators rate the remaining plots on a five-point Likert scale (1–5) for aesthetics and information richness, and confirm the absence of sensitive content. Only plots scoring $\geq$ 4 on both criteria and passing manual verification are retained. This hybrid process ensures high visual fidelity, semantic clarity, and ethical compliance throughout PlotGen-Bench.

\vpara{Ground-truth Code Annotation.} 
To construct a reliable ground-truth code for each plot, we follow a human-centric “annotate-and-verify” workflow. For each filtered image, one expert writes executable code that accurately reproduces the visualization’s layout, style, and data representation, while another independently reviews it for correctness and fidelity. Discrepancies are resolved through discussion or senior expert consultation. This two-stage process ensures all PlotGen-Bench code samples are executable, visually consistent, and semantically faithful to their reference plots.

\vpara{Plot Translation Dataset Construct.} 
Beyond the primary replication and multi-library datasets, PlotGen-Bench includes a plot-to-plot translation subset for evaluating transformation tasks. Human experts first examine 29 plot types (excluding animated plots) to identify 27 valid conversion pairs that preserve data semantics (e.g., bar → line, pie → donut). For each pair, structured prompts ( See Appendix~\ref{sec:translation-prompt-appendix}) guide GPT-5 in generating coherent datasets visualizable in both forms. The resulting pairs (See Appendix~\ref{sec:generated_paris}) are then expert-reviewed for semantic consistency, visual plausibility, and transformation accuracy. After multi-stage validation, 27 high-quality translation instances, each with shared data and verified ground-truth code—are retained.

\begin{figure*}[t!]
    \centering
    \begin{subfigure}[b]{0.6\linewidth}
        \centering
        \includegraphics[width=\textwidth]{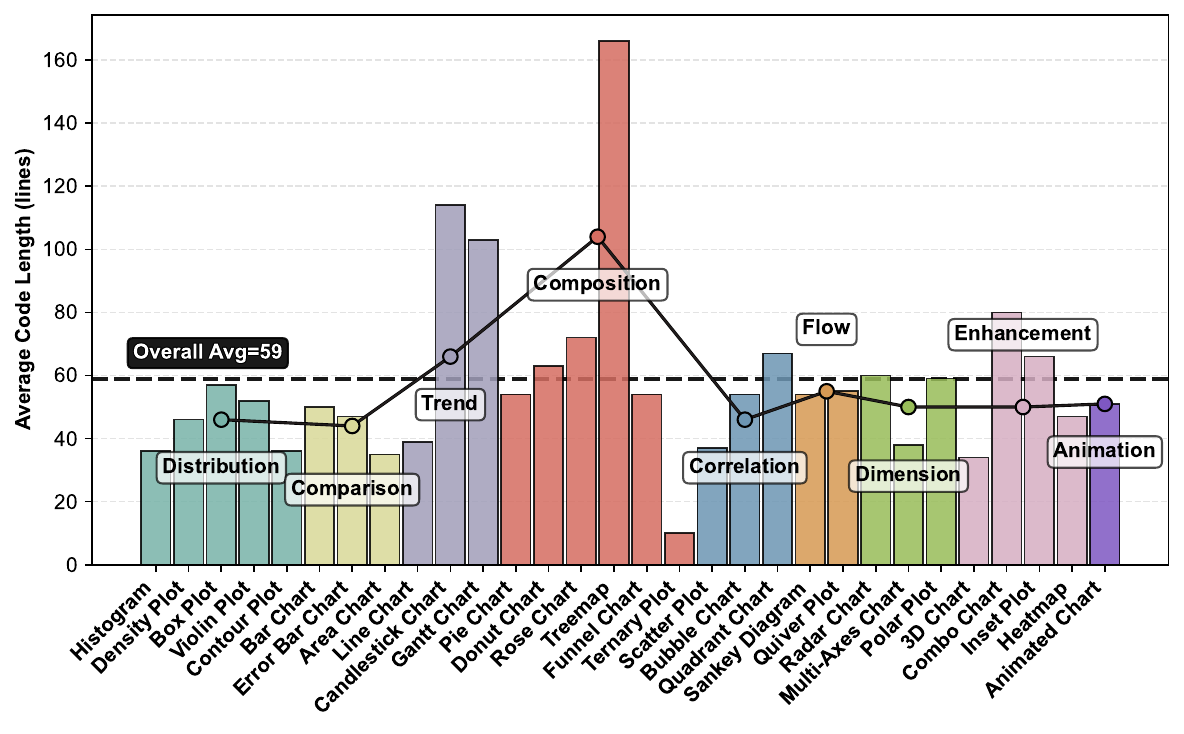}
        \caption{Distribution of code line length}
        \label{fig:code_length}
    \end{subfigure}
    \hfill
    \begin{subfigure}[b]{0.35\linewidth}
        \centering
        \includegraphics[width=\linewidth]{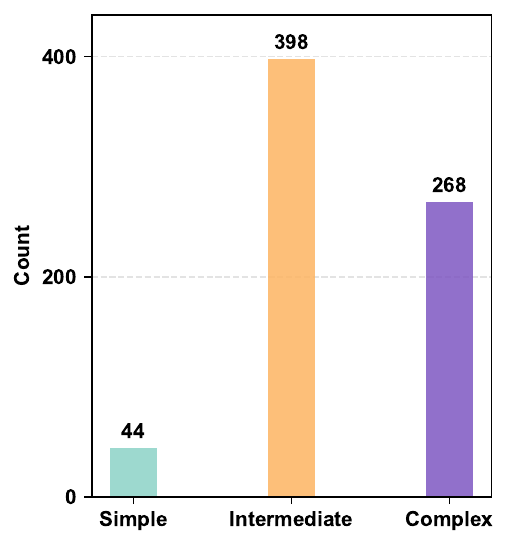}
        \caption{Distribution of visual complexity.}
        \label{fig:figure_comple}
    \end{subfigure}
    \caption{Complexity Analysis of Code and Plots} 
    \label{fig:erros}
    \vspace{-5mm}
\end{figure*}
\subsection{Data Analysis}\label{sec:data_analysis}
\vpara{Diversity of Visualization Coverage.} 
A central objective of PlotGen-Bench is to provide broad, representative coverage of visualization types and libraries. The benchmark includes 710 curated plots across 30 types grouped into 9 analytical categories, spanning conventional 2D plots (e.g., bar, line, scatter), advanced forms (e.g., treemap, radar, heatmap), and complex formats like 3D, combo, and animated plots.
As shown in Figure~\ref{fig:plot_distribution}, while standard 2D plots remain well represented, greater emphasis is placed on complex forms that test plot-to-code generalization. The Enhancement category dominates with over 28\% of samples, driven by 3D and combo plots requiring multi-axis reasoning and layered rendering. Composition-based visualizations (e.g., treemap, donut) form another major cluster, emphasizing hierarchical and proportional structures. About 5\% of samples are animated plots, introducing temporal and rendering challenges absent in prior benchmarks. Detailed taxonomy and representative examples appear in Appendix~\ref{sec:detaild category}.

\vpara{Complexity and Difficulty Analysis.} 
To assess the challenges posed by PlotGen-Bench, we analyze the intrinsic complexity of its plots and ground-truth code.
As shown in Figure~\ref{fig:code_length}, the dataset’s average code length is 59 lines, with large variation across categories. Even typically simple plots (e.g., pie, donut, funnel) exhibit high complexity, often matching or exceeding the overall average.
Using GPT-5, we further stratify samples into three difficulty tiers—simple, intermediate, and complex—based on visual and structural attributes (See Appendix~\ref{sec:evalution_complexity_prompt}). As illustrated in Figure~\ref{fig:figure_comple}, 37\% of plots are complex, and over 93\% are non-simple, underscoring the benchmark’s focus on realistic, compositionally rich, and technically demanding visualization scenarios.
This stratified design supports fine-grained evaluation of model robustness across varying visual and code-generation complexities.

\vpara{Multi-Library Distribution.}
Beyond overall complexity, PlotGen-Bench exhibits strong library-level diversity, mirroring the heterogeneity of modern Python visualization ecosystems. It spans five major libraries—Matplotlib, Seaborn, Plotly, NetworkX, and Plotnine—covering both static and interactive paradigms. Among them, Matplotlib accounts for 23.5\%, Seaborn 25.5\%, Plotly 23\%, NetworkX 18.4\%, and Plotnine 9.7\%. Detailed comparisons across libraries are provided in Appendix~\ref{sec:details_about_librabies}.

\begin{figure*}
    \centering
    \includegraphics[width=\linewidth]{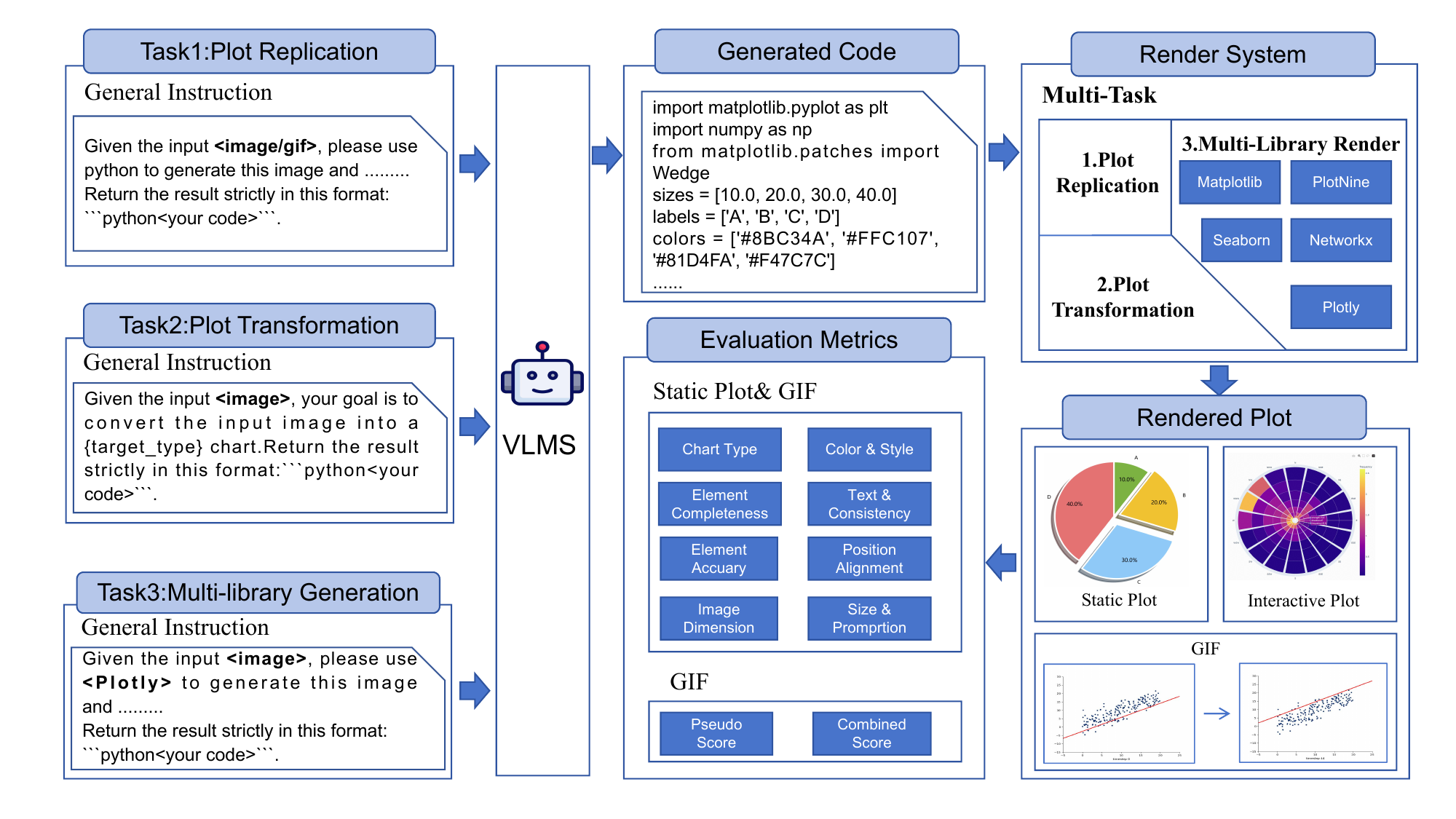}
    \caption{The pipeline of PlotGen-Bench. PlotGen-Bench supports 5 Python rendering libraries, 3 tasks, and rendering validation across 30 types of visualizations, including complex plots such as 3D plots, animations, and interactive plots.}
    \label{fig:eval_pipeline}
\end{figure*}
\subsection{Discussion}
A distinguishing feature of PlotGen-Bench is its deliberate inclusion of rarely studied visualization types, such as Quadrant Plots, Gantt Charts, Polar Plots, Sankey Diagrams, and Animated Plots that are absent from prior benchmarks. Notably, Sankey and Animated plots pose significant challenges due to their hierarchical flows and temporal dynamics, which strain current VLMs’ ability to model multi-stage rendering and maintain temporal consistency. By integrating these complex families, PlotGen-Bench extends plot-to-code evaluation beyond conventional static 2D reproduction.

Moreover,  the data collection process prioritizes information richness, stylistic diversity, and compositional complexity, ensuring each visualization reflects authentic real-world design rather than simplified templates. This contrasts with earlier, template-driven benchmarks. The combination of visual diversity and structural difficulty creates a realistic and comprehensive testbed, requiring models to generate not only syntactically correct code but also robust, semantically coherent, and stylistically adaptive visualizations.

\section{Evaluation Metrics}

To assess the performance of VLMs on PlotGen-Bench, we design a hybrid evaluation protocol that combines automated execution checks, VLM-based scoring, and specialized dynamic analysis for animated visualizations, as illustrated in Figure~\ref{fig:eval_pipeline}. You can find plot generation instruction examples in Appendix \ref{sec:instruction_exmaples_differnet_tasks}
and evaluation case  study in Appendix \ref{sec:evalution_case_study_gpt4o}.

\vpara{Code Execution Pass Rate.} The Code Execution Pass Rate (PR) quantifies the executability of model-generated code by measuring the proportion of samples that successfully run in a controlled Python environment without triggering syntax or runtime errors.

\begin{table*}[t!]
\centering
\vspace{-2mm}
\caption{Performance comparison  (0–100) of vision–language models on PlotGen-Bench across different plot types. Overall scores are computed as macro-averages across all sub-categories, excluding Animation for fair comparison.}
\resizebox{\textwidth}{!}{%
\setlength{\tabcolsep}{1mm}{
\begin{tabular}{lcccccccccc}
\toprule
Model & Distribution & Comparison & Trend & Composition & Correlation & Flow & Dimension & Enhancement & Animation & Overall\\
\midrule
\multicolumn{11}{c}{\textbf{Open-source VLM}} \\
\midrule
InternVL3-9B &16.0 &26.8 &21.7 &20.8 &13.3 &25.0 &17.7 &25.3 &2.5 &21.8 \\
InternVL3-78B &43.4 &55.7 &39.6 &47.3 &49.0 &40.4 &55.9 &48.0 &33.2 &48.3 \\
Qwen2.5-VL-7B-Instruct &34.3 &47.2 &28.0 &31.7 &28.6 &25.4 &44.4 &35.6 &2.6 &35.2 \\
Qwen2.5-VL-72B-Instruct &56.2 &61.1 &48.0 &52.8&60.2&51.7 &\textbf{71.8} &55.4 &32.0 &56.4 \\
MiMo-VL-7B-SFT &22.5 &29.0 &24.3 &25.4 &28.9 &23.7 &28.7 &31.2 &7.0 &27.4 \\
MiMo-VL-7B-RL &21.2 &24.0 &25.5 &23.4 &29.2 &18.1 &24.8 &28.9 &6.4 &25.7 \\
Kimi-VL-A3B-Instruct &33.4 &39.9 &40.6 &25.5 &31.0 &31.0 &30.3 &38.0 &17.7 &34.4 \\
Kimi-VL-A3B-Thinking &18.3 &34.4 &17.3 &16.6 &19.5 &10.8 &19.2 &21.0 &5.5 &20.3 \\
GLM-4.1V-9B-Thinking &32.4 &44.2 &31.9 &33.8 &35.0 &47.0 &34.4 &40.0 &-- &26.9 \\
GLM-4.5V &48.6 &52.0 &44.2 &43.5 &40.2 &28.9 &51.2 &50.9 &-- &47.2 \\
Qwen3-VL-235B-A22B-Instruct &\textbf{64.7} &\textbf{71.0} &\textbf{60.6} &\textbf{66.4} &\textbf{66.8} &\textbf{61.5} &65.1 &\textbf{62.3} &\textbf{38.1} &\textbf{64.6} \\
\midrule
\multicolumn{11}{c}{\textbf{Closed-source VLM}} \\
\midrule
Claude-Sonnet-4-20250514-Thinking &76.7 &76.2 &72.7 &74.3 &76.1 &50.1 &\textbf{82.9} &72.5 &44.5 &74.1 \\
Claude-3.7-Sonnet-20250219-Thinking &73.7 &79.9 &72.8 &74.4 &\textbf{81.0} &69.5 &67.6 &72.8 &34.5 &74.0 \\
GPT-5-2025-08-07 &\textbf{78.7} &\textbf{87.8} &\textbf{76.1} &\textbf{78.3} &78.4 &\textbf{75.0} &81.9 &\textbf{79.9} &\textbf{45.2} &\textbf{79.7} \\
GPT-4o-2024-11-20 &69.1 &71.5 &69.1 &58.5 &68.2 &68.7 &70.1 &69.8 &41.7 &67.7 \\
GPT-4o-Mini-2024-07-18 &51.2 &62.8 &54.4 &52.4 &51.8 &54.9 &61.2 &57.2 &37.1 &55.5 \\
Gemini-2.5-Pro &73.1 &76.8 &70.7 &76.0 &75.3 &59.5 &76.3 &67.7 &-- &72.0 \\
Gemini-2.5-Flash &57.3 &60.3 &44.7 &55.1 &61.9 &54.1 &56.0 &59.9 &-- &60.0 \\
Doubao-1.5-Thinking-Pro-Vision-250415 &48.6 &55.7 &52.0 &38.6 &48.9 &45.6 &51.8 &50.8 &28.6 &48.7 \\
Doubao-Seed-1-6-Thinking-250715 &55.2 &67.9 &54.0 &54.5 &52.7 &47.7 &57.3 &58.9 &32.5 &57.1 \\
\bottomrule
\end{tabular}
}}
\label{tab:results_on_diff_types}
\vspace{-2mm}
\end{table*}

\vpara{VLM-based Scoring.} 
Beyond code execution pass rate, we use GPT-4o as an automated evaluator to compute a visual score (GS) comparing model-rendered plots with ground-truth visualizations. Acting as a domain expert under structured prompts (See Appendix~\ref{sec:evaluation_prompt}), GPT-4o assesses seven dimensions, including \textit{chart type accuracy}, \textit{element completeness}, \textit{positional alignment}, \textit{proportional scaling}, \textit{color and style consistency}, \textit{textual correspondence}, and \textit{overall structural coherence}, each rated from 0–100 with brief justifications. 

\vpara{GIF Dynamics Evaluation.} 
For animated GIFs, we introduce a temporal consistency evaluation based on frame-level perceptual similarity. First, GPT-4o assesses visual similarity between the first and last frames. Then, to detect pseudo-animations (lacking real motion), we sample multiple frames and compute their perceptual hashes (pHash) to measure low-frequency visual differences. GIFs with small hash distances are penalized to reflect limited temporal dynamics. Implementation details are provided in Appendix~\ref{sec:animation_evalution}.

\section{Experiments}
\begin{table*}[h]
\centering
\vspace{-1mm}
\vspace{-2mm}
\label{tab:main_results}
\caption{Performance comparison of vision–language models on PlotGen-Bench across different libraries. PR denotes the pass rate (\%), and GS denotes the GPT-4o score (0–100).}
\setlength{\tabcolsep}{2.0mm}{
\footnotesize
\begin{tabular}{lcccccccccc}
\toprule
 \multirow{2}{*}{Model} & \multicolumn{2}{c}{Matplotlib}  & \multicolumn{2}{c}{Seaborn} & \multicolumn{2}{c}{Plotly} & \multicolumn{2}{c}{Plotnine} & \multicolumn{2}{c}{NetworkX}\\
 \cmidrule(l{0.5em}r{0.5em}){2-3} \cmidrule(l{0.5em}r{0.5em}){4-5} \cmidrule(l{0.5em}r{0.5em}){6-7} \cmidrule(l{0.5em}r{0.5em}){8-9} \cmidrule(l{0.5em}r{0.5em}){10-11}  
 & PR & GS & PR & GS  & PR & GS  & PR & GS  & PR & GS\\
\midrule
\multicolumn{11}{c}{\textbf{Open-source VLM}} \\
\midrule
InternVL3-9B   &43.4 &20.7 &20.4 &6.9 &46.5 &19.3 &0.0 &0.0 &41.7 &21.5 \\
InternVL3-78B  &71.7 &43.1 &69.4 &42.7 &\textbf{81.4} &52.0 &21.1 &11.9 &86.1 &56.5 \\
Qwen2.5-VL-7B-Instruct   &71.7 &28.7 &57.1 &31.3 &79.1 &45.0 &0.0 &0.0 &75.0 &35.2 \\
Qwen2.5-VL-72B-Instruct  &84.7 &53.1 &77.6 &55.1 &79.1 &53.3 &15.8 &13.5 &83.3 &58.0 \\
MiMo-VL-7B-SFT &56.5 &30.7 &34.7 &17.2 &2.3 &1.0 &10.5 &2.7 &47.2 &21.8 \\
MiMo-VL-7B-RL &60.9 &27.4 &28.6 &16.2 &7.0 &4.2 &0.0 &0.0 &47.2 &26.2 \\
Kimi-VL-A3B-Instruct &32.6 &8.8 &61.2 &34.0 &18.6 &12.4 &36.8 &15.8 &72.2 &35.1 \\
Kimi-VL-A3B-Thinking &45.7 &12.0 &26.5 &14.4 &11.6 &6.3 &0.0 &0.0 &38.9 &18.2 \\
GLM-4.1V-9B-Thinking &58.7 &34.1 &46.9 &33.0 &51.1 &32.5 &0.0 &0.0 &61.1 &43.1 \\
GLM-4.5V &58.7 &44.4 &49.0 &38.7 &32.5 &22.4 &21.1 &17.3&52.8 &39.5 \\
Qwen3-VL-235B-A22B-Instruct &\textbf{93.5} &\textbf{62} &\textbf{87.8} & \textbf{62.8}&79.1 &\textbf{58.1} &\textbf{63.2} &\textbf{41.6} & \textbf{97.2}& \textbf{72.0}\\
\midrule
\multicolumn{11}{c}{\textbf{Closed-source VLM}} \\
\midrule
Claude-Sonnet-4-20250514-Thinking &\textbf{97.8} &74.4 &\textbf{93.9} &75.7 &83.7 &59.3 &36.8 &29.5 &\textbf{100.0} &81.1 \\
Claude-3.7-Sonnet-20250219-Thinking &93.5 &78.6 &87.8 &73.0 &87.0 &64.2 &5.3 &4.2 &94.4 &79.8 \\
GPT-5-2025-08-07 &93.5 &\textbf{82.7} &91.8 &78.9 &\textbf{88.3} &\textbf{76.9} &36.8 &27.7 &86.1 &72.1 \\
GPT-4o-2024-11-20 &89.1 &65.6 &85.7 &64.4 &86.0 &61.5 &\textbf{68.4} &\textbf{41.3} &91.7 &69.5 \\
GPT-4o-Mini-2024-07-18 &82.6 &52.2 &75.5 &54.0 &79.1 &49.0 &42.1 &28.6 &91.7 &52.2 \\
Gemini-2.5-Pro &82.6 &74.2 &89.8 &\textbf{80.9} &72.1 &60.4 &10.5 &8.0 &97.2 &\textbf{85.0} \\
Gemini-2.5-Flash &67.3 &56.2 &77.5 &67.1 &44.2 &37.2 &15.7 &13.0 &72.2 &63.4 \\
Doubao-1.5-Thinking-Pro-Vision-250415 &65.2 &50.1 &36.7 &28.0 &44.2 &34.2 &10.5 &8.0 &61.1 &46.8 \\
Doubao-Seed-1-6-Thinking-250715 &80.4 &65.8 &53.1 &40.4 &34.8 &30.0 &26.3 &20.8 &66.7 &54.8 \\
\bottomrule
\end{tabular}
\label{tab:results_on_diff_libs}
}
\vspace{-4mm}
\end{table*}

\subsection{Evaluation Models}
In our benchmark evaluation, we assess a representative set of state-of-the-art VLMs, including both open-source models (such as InternVL3 \cite{wang2025internvl3}, Qwen3-VL \cite{yang2025qwen3}, GlM4.5v \cite{vteam2025glm45vglm41vthinkingversatilemultimodal}, MiMo-VL \cite{xiaomi2025mimo}, Kimi-VL \cite{team2025kimi}) and closed-source models (such as the GPT series \cite{openai2024gpt4technicalreport}, Claude series \cite{anthropic2025claude4}, Gemini series \cite{huang2025gemini}, Doubao\cite{guo2025seed1}). The models span parameter scales from 7B to 235B, enabling a comprehensive and representative comparison of their capabilities across plot replication, plot transformation, and multi-library generation tasks.

\subsection{Results on Difference Plot Types}
Table~\ref{tab:results_on_diff_types} presents VLM performance on PlotGen-Bench across plot types using the GPT-Score metric, with pass-rate (PR) results in Appendix~\ref{sec:passing_rate_of_models}. Closed-source models consistently outperform open-source ones but show large variance across categories. GPT-5, Claude-Sonnet-4, and Claude-3.7-Sonnet excel on simpler plots (distribution, comparison, trend) yet drop sharply on complex types like animation, where styling, interactivity, and temporal logic challenge code generation. This indicates that while current VLMs can reproduce basic structures, they often fail to maintain precise layouts, color mappings, or animation behavior. Among open-source models, Qwen3-VL-235B-A22B-Instruct, Qwen2.5-VL-72B, and InternVL3-78B perform best, suggesting larger capacity aids plot-to-code generation. Overall, progress remains limited to static, structurally regular plots, whereas dynamic and stylistically rich visualizations demand better layout-aware pretraining, explicit plot-grammar supervision, and execution-level feedback. Detailed statistics can be found in Appendix~\ref{sec:each_type_PR_GS}.

\subsection{Results on Different Libraries}
Table~\ref{tab:results_on_diff_libs} presents the performance comparison of various VLMs on PlotGen-Bench across five popular plotting libraries, evaluated by both PR and GS. The results reveal large discrepancies across libraries, reflecting each model’s ability to generalize plot-to-code generation beyond a single syntax or rendering paradigm.
Closed-source models, particularly GPT-5, Claude-Sonnet-4, and Claude-3.7-Sonnet, exhibit strong robustness and consistent performance across diverse visualization libraries, including \textit{Matplotlib}, \textit{Seaborn}, \textit{Plotly}, and \textit{NetworkX}.
This indicates that these models effectively capture abstract plotting semantics rather than memorizing syntax-specific templates. However, performance drops sharply on \textit{Plotnine}, which adopts a declarative grammar-of-graphics paradigm that differs fundamentally from the procedural plotting logic dominant in other Python libraries. These results highlight a persisting gap in multi-library adaptability—current models tend to overfit to the dominant Matplotlib/Seaborn ecosystem.
Among open-source VLMs, it is worth noting that Qwen3-Vl-235B-A22B-Instruct has achieved results comparable to those of some closed-source models. While Qwen2.5-VL-72B and InternVL3-78B achieve competitive results on \textit{Matplotlib} and \textit{Seaborn}, showing their ability to generate structurally valid and partially executable code. Nonetheless, their performance collapses on \textit{Plotnine}, underscoring a limited capacity for multi-library generalization.
Overall, these findings reveal that while current VLMs have achieved notable progress in handling mainstream plotting libraries, their ability to generalize plot-to-code generation across multiple libraries remains limited.

\subsection{Results on Plot Transformation}
Table~\ref{tab:results_on_plot_trans} presents the results of the Plot Transformation task, which measures a model’s ability to modify existing visualizations into new ones while maintaining semantic coherence and executable correctness. We observe that most VLMs achieve strong execution rates, yet their GPT-scores remain notably lower, indicating that while the generated code is often syntactically correct and executable, the resulting plots frequently deviate from the intended visual semantics, revealing a gap between code-level accuracy and visual-level fidelity in plot-to-plot transformation.
This discrepancy highlights a persistent gap between code-level accuracy and visual-level fidelity, underscoring the challenge of achieving semantically consistent plot-to-plot transformations.

\begin{table}[t!]
    \centering
    \footnotesize
    \caption{Performance comparison of vision–language models on Plot Transformation. PR denotes the pass rate (\%), and GS denotes the GPT-4o score (0–100). }
    \begin{tabular}{c|c|c}
    
    \toprule
      Model   & PR & GS \\
      \midrule
       GLM-4.5V  &85.2  &50.4 \\
       Qwen3-VL-235B-A22B-Instruct &\textbf{100} &56.4\\
       \midrule
       GPT-5-2025-08-07 & 96.3& 59.1\\
       Gemini-2.5-Pro  & \textbf{100} & 60.2 \\
       Claude-Sonnet-4-20250514-Thinking & \textbf{100} &\textbf{63.0}  \\
       Doubao-Seed-1-6-Thinking-250715 &88.9&50.4  \\
    \bottomrule
    \end{tabular}
    \vspace{-4mm}
    \label{tab:results_on_plot_trans}
\end{table}

\subsection{Different Complexity Levels}
\begin{figure}
    \centering
    \includegraphics[width=0.8\linewidth]{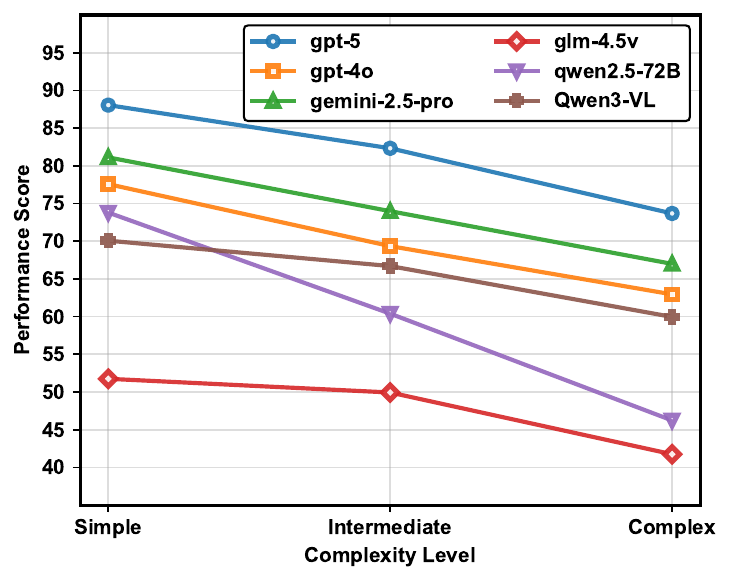}
    \caption{GPT-score of models at different complexity levels}
    \label{fig:model_performance_complexity}
    \vspace{-6mm}
\end{figure}
We further report model performance across different levels of plot complexity.
As shown in Figure \ref{fig:model_performance_complexity}, model performance consistently declines as the visual complexity of plots increases.
Notably, several models achieve lower scores on the simple level than GPT-5 does on the intermediate level, highlighting the superior capability of GPT-5 in plot code generation.
Moreover, we observe that open-source models often perform worse on simple tasks than closed-source models do on complex ones, indicating that open-source systems still lag behind in handling sophisticated visual reasoning and generation tasks, and require substantial improvement to bridge this gap.

\section{Additional Experiments}
Specific error analysis cases can be found in Appendix \ref{sec:error-analysis}. Classifications of errors are provided in the Appendix \ref{sec:code-related-error-definition},\ref{sec:rendering-related-error-definition}. The analysis of Pseudo-animation probabilities of different models is provided in the Appendix \ref{sec:pseudo_proba}.
\section{Conclusion}
In this work, we present PlotGen-Bench, a benchmark for evaluating the code generation abilities of vision-language models (VLMs) under realistic and complex visualization requirements. It covers 9 categories, 30 subcategories, and 3 tasks—plot replication, transformation, and multi-library generation—across five major libraries: Matplotlib, Seaborn, Plotly, Plotnine, and NetworkX.
Experiments reveal that current VLMs achieve high executability but struggle to maintain visual fidelity, particularly in transformation and cross-library scenarios. Closed-source models perform better on mainstream libraries, while some open-source ones excel on niche types.
These findings highlight the need for functionally grounded multimodal learning that integrates execution feedback, multi-library alignment, and visual supervision.
We hope that PlotGen-Bench will serve as a catalyst for developing next-generation VLMs capable of reliable, interpretable, and semantically consistent visualization generation.

\section*{Limitations}
Although this benchmark employs multiple strategies to preserve the informativeness, complexity, and diversity of the images, potential biases may still arise due to implicit preferences of human annotators or curators. Moreover, future work could extend beyond simple chart-type transformations to include more sophisticated and nuanced editing tasks, such as structural modifications, multi-step reasoning, or style-preserving content adaptation to better reflect the full spectrum of real-world visualization challenges. 

\bibliography{custom}
\newpage
\appendix

\section{Comparative Analysis of Code Execution Pass Rates (PR) of Different Models}
\label{sec:passing_rate_of_models}
\begin{table*}[h]
\centering
\vspace{-2mm}
\renewcommand{\arraystretch}{1.5}
\caption{Code pass-rate (0-100) comparison of vision–language models on PlotGen-Bench across different plot types. Overall scores are computed as macro-averages across all sub-categories, excluding Animation for fair comparison. }
\resizebox{\textwidth}{!}{%
\setlength{\tabcolsep}{1mm}{
\begin{tabular}{lcccccccccc}
\toprule
Model & Distribution & Comparison & Trend & Composition & Correlation & Flow & Dimension & Enhancement & Animation & Overall\\
\midrule
\multicolumn{11}{c}{\textbf{Open-source VLM}} \\
\midrule
InternVL3-9B   &36.3 &43.9 &41.3 &44.1 &30 &54.5 &34.7 &50.2 &22.9 &43.1 \\
InternVL3-78B   &73.5 &83.3 &60.3 &83.1 &76 &81.8 &81.6 &75.6 &\textbf{88.6} &76.6 \\
Qwen2.5-VL-7B-Instruct   &66.7 &75.8 &57.1 &77.1 &62 &59.1 &71.4 &66.3 &17.1 &68.1 \\
Qwen2.5-VL-72B-Instruct   &77.5 &78.8 &74.6 &83.1&86&90.9&\textbf{98} &82.4 &82.9 &82.4 \\
MiMo-VL-7B-SFT    &41.2 &42.4 &39.7 &44.1 &48 &40.9 &42.9 &49.3 &25.7 &44.7 \\
MiMo-VL-7B-RL  &42.2 &37.9 &41.3 &44.9 &50 &40.9 &40.8 &48.3 &25.7 &44.4 \\
Kimi-VL-A3B-Instruct   &58.8 &60.6 &73 &53.4 &60 &63.6 &49 &62 &54.3 &59.9 \\
Kimi-VL-A3B-Thinking  &36.3 &53 &33.3 &36.4 &38 &36.4 &34.7 &41.5 &25.7 &39.3 \\
GLM-4.1V-9B-Thinking  &46.1 &56.1 &46 &54.2 &52 &72.7 &49 &58.5 &- &53.8 \\
GLM-4.5V &60.8 &63.6 &57.1 &58.5 &54 &54.5 &61.2 &69.8 &- &62.4 \\
Qwen3-VL-235B-A22B-Instruct &\textbf{88.2}&\textbf{87.9}&\textbf{84.1}&\textbf{94.1}&\textbf{92.0}&\textbf{95.5}&85.7&\textbf{88.8}&\textbf{88.6} &\textbf{89.3}\\
\midrule
\multicolumn{11}{c}{\textbf{Closed-source VLM}} \\
\midrule
Claude-Sonnet-4-20250514-Thinking     &\textbf{94.1} &89.4 &92.1 &\textbf{95.8} &94 &72.7 &\textbf{98} &\textbf{93.7} &85.7 &\textbf{93.2} \\
Claude-3-7-Sonnet-20250219-Thinking   &87.3 &90.9 &85.7 &93.2 &\textbf{96} &\textbf{95.5} &81.6 &89.8 &85.7 &89.8 \\
GPT-5-2025-08-07   &89.2 &\textbf{97} &87.3 &89 &90 &86.4 &93.9 &91.7 &\textbf{97.1} &90.8 \\
GPT-4o-2024-11-20     &91.2 &89.4 &\textbf{93.7} &83.1 &90 &\textbf{95.5} &89.8 &92.7 &80.0 &90.2 \\
GPT-4o-Mini-2024-07-18   &76.5 &84.8 &81 &87.3 &74 &90.9 &87.8 &83.4 &71.4 &82.8 \\
Gemini-2.5-Pro   &83.3 &83.3 &77.8 &87.3 &86 &72.7 &85.7 &78.5 &- &82.1 \\
Gemini-2.5-Flash       &66.7 &66.7 &50.8 &63.6 &72 &68.2 &63.3 &72.2 &- &66.5 \\
Doubao-1.5-Thinking-Pro-Vision-250415  &62.7 &68.2 &65.1 &51.7 &62 &63.6 &65.3 &67.3 &60.0 &63.1 \\
Doubao-Seed-1-6-Thinking-250715    &70.6 &81.8 &69.8 &67.8 &68 &68.2 &71.4 &76.1 &71.4 &72.6 \\
\bottomrule
\end{tabular}
\label{tab:passing_rate_on_diff_types}
}}
\end{table*}
As shown in Table \ref{tab:passing_rate_on_diff_types}, larger models, particularly in the open-source group, demonstrate stronger syntactic stability and execution robustness, with higher pass rates reflecting their ability to internalize plotting syntax and reduce runtime errors. However, this improvement primarily affects code-level fluency rather than visual semantic accuracy. Second, a template-driven generation tendency is evident: many models, both open- and closed-source, achieve high execution success yet produce visually imperfect plots, indicating a reliance on safe, memorized code structures rather than context-aware, semantically faithful plotting. Finally, while open-source models have largely caught up to closed-source systems in execution rates, a substantial gap in visual grounding remains. Closed-source models benefit from execution-aware fine-tuning and human-in-the-loop corrections, resulting in superior alignment between code and intended visual output. Overall, scaling enhances robustness, template-driven strategies constrain visual fidelity, and semantic grounding remains the key challenge for achieving parity across model types.

\section{Error Analysis}
\label{sec:error-analysis}
We perform a comprehensive error analysis of the state-of-the-art GPT-5 model to identify potential avenues for improving VLMs. Our investigation categorizes errors into two primary types: code-related errors and rendering-related errors.

\begin{figure}[htbp]
    \centering
    \begin{subfigure}[b]{0.23\textwidth}
        \centering
        \includegraphics[width=\textwidth]{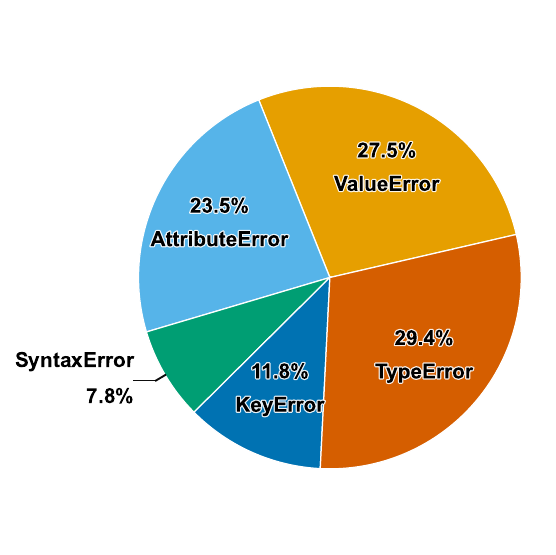}
        \caption{Distribution of CR}
        \label{fig:code-related-errors-ds}
    \end{subfigure}
    \hfill
    \begin{subfigure}[b]{0.23\textwidth}
        \centering
        \includegraphics[width=\textwidth]{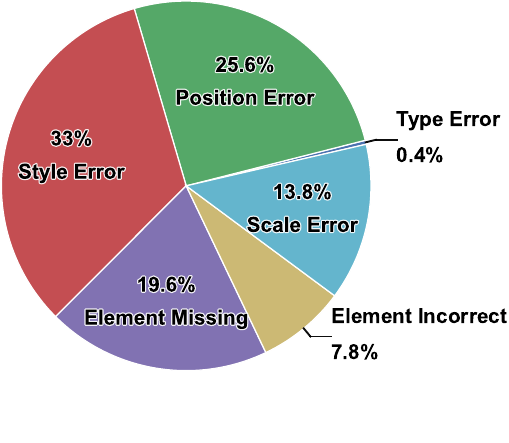}
        \caption{Distribution of RE}
        \label{fig:render-related errors-ds}
    \end{subfigure}
    \caption{Error analysis of GPT5 across two type of errors.}
    \label{fig:erros}
\end{figure}
\vpara{Code-related errors (CR).}  
As shown in Figure \ref{fig:code-related-errors-ds}, we categorize exceptions raised during code execution into five types: SyntaxError, TypeError, AttributeError, ValueError, and KeyError/IndexError. Detailed definitions of these categories and corresponding examples can be found in Appendix \ref{sec:code-related-error-definition}. From Figure \ref{fig:code-related-errors-ds}, it is evident that the predominant errors in GPT-5-generated code are TypeError, ValueError, and AttributeError. The first two indicate that the model can correctly recall APIs but often makes mistakes in numerical types or size inference, whereas the latter suggests that the model frequently calls non-existent methods, reflecting either memory errors or the use of outdated or deprecated APIs.

\vpara{Rendering-related errors (RE).}
As shown in Figure \ref{fig:render-related errors-ds}, we categorize rendering-related errors into six types: Chart Type Error (CTE), Position Alignment Error (PAE), Color \& Style Error (CSE), Scale \& Size Error (SSE), Element Missing Error (EME), and Element Incorrect Error (EIE). Definitions of these categories and corresponding examples can be found in Appendix \ref{sec:rendering-related-error-definition}. We observe that the primary issues in the images generated by large models lie in color style, element positioning, and missing elements, indicating that there is still substantial room for improvement in aligning visual information with language. Particularly for GIF generation, we find that even the strongest GPT-5 model has nearly a 60\% probability of producing pseudo-animated images. 

\section{Pseudo-animation probabilities of different models}
\label{sec:pseudo_proba}
\begin{table}[]
    \centering
    \footnotesize
    \renewcommand{\arraystretch}{1.5}
    \caption{True   -animation probabilities of
different models}
    \resizebox{0.45\textwidth}{!}{%
    \begin{tabular}{c|c}
    
    \toprule
      Model   & True-animation Probability\\
      \midrule
      \multicolumn{2}{c}{\textbf{Open-source VLM}} \\
      \midrule
       InternVL3-9B  &5.7   \\
       InternVL3-78B &37.1 \\
       Qwen2.5-VL-7B-Instruct&5.7\\
       Qwen2.5-VL-72B-Instruct&\textbf{57.1}\\
       MiMo-VL-7B-SFT &2.9\\
       MiMo-VL-7B-RL&2.9\\
       Kimi-VL-A3B-Instruct&17.6\\
       Kimi-VL-A3B-Thinking&5.7\\
       GLM-4.1V-9B-Thinking&--\\
       GLM-4.5V&--\\
       Qwen3-VL-235B-A22B-Instruct&54.3\\
       \midrule
        \multicolumn{2}{c}{\textbf{Closed-source VLM}} \\
       \midrule
       Claude-Sonnet-4-20250514-Thinking &\textbf{74.3}\\
       Claude-3.7-Sonnet-20250219-Thinking&40.0\\
       GPT-5-2025-08-07 & 60.0\\
       GPT-4o-2024-11-20&\textbf{74.3}\\
       GPT-4o-Mini-2024-07-18&68.6\\
       Gemini-2.5-Pro  & -- \\
       Gemini-2.5-Flash&--\\
       Doubao-1.5-Thinking-Pro-Vision-250415 & 51.4  \\
       Doubao-Seed-1-6-Thinking-250715 &34.3 \\
    \bottomrule
    \end{tabular}
    \label{tab:true_proba_of_models}
    }
    
\end{table}
Table \ref{tab:true_proba_of_models} presents the pseudo-animation probabilities of different VLMs, reflecting how often the generated GIFs contain true frame-wise motion rather than static images mislabeled as GIFs. Overall, most open-source models exhibit a high rate of pseudo-animations, indicating limited understanding of temporal visualization semantics. Among them, Qwen2.5-VL-72B-Instruct achieves the best performance (57.1\%), while smaller or instruction-tuned variants such as MiMo-VL-7B and Kimi-VL-A3B-Thinking almost entirely fail to produce genuine motion (< 10\%). In contrast, closed-source VLMs demonstrate substantially stronger temporal generation capabilities: Claude-Sonnet-4 (74.3\%) and GPT-4o (74.3\%) lead the ranking, followed by GPT-4o-Mini (68.6\%) and GPT-5 (60\%). These results suggest that while open-source VLMs can render syntactically correct GIFs, they often lack dynamic reasoning and temporal control, whereas advanced closed-source models achieve more consistent and realistic animation synthesis through stronger multimodal alignment and instruction comprehension.

\section{The Detailed Statistical Information about Each Type of Plot}
\begin{figure*}
    \centering
    \includegraphics[width=1\linewidth]{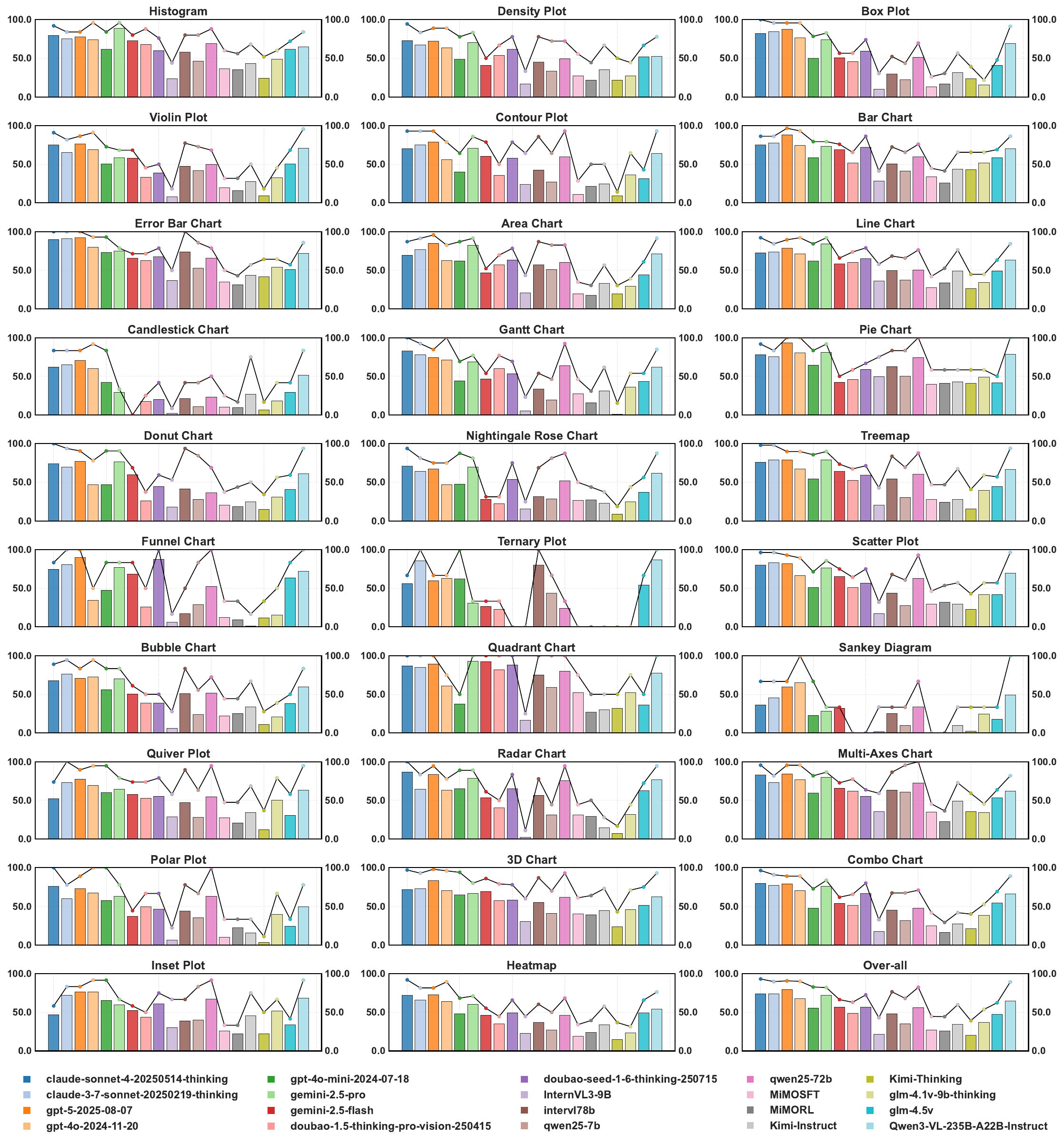}
    \caption{Performance analysis of different models across all types of plots, in terms of PR (Pass Rate) and GS (GPT Score). In the chart, the bars represent GPT scores, and the lines represent code pass rates.}
    \label{fig:PR_GS_each_type}
\end{figure*}

Here, we present the Pass Rate (PR) and GPT-based Score (GS) of 20 models across 29 plot types (The result of animated plot has reported in the paper). As shown in the Figure \ref{fig:PR_GS_each_type}, for relatively simple plots such as bar, line, and pie charts, most models achieve a high code executability rate, yet their scores vary significantly. This discrepancy arises because the data collected in PlotGenBench exhibit high stylistic diversity and dense information content, making it challenging for open-source or small-parameter models to perform complex reasoning, which consequently leads to lower scores.

Furthermore, we observe that, except for the Sankey diagram, closed-source models generally achieve satisfactory levels in both score and code execution rate. In contrast, lightweight open-source models struggle when dealing with complex visualizations, indicating a pressing need for improvement.

\label{sec:each_type_PR_GS}

\section{Prompt Template for Synthesizing Plots Transformation Subset}
\label{sec:translation-prompt-appendix}
The “Plot Transformation Synthesis Prompt” (See Figure \ref{fig:sytheic}) is designed to standardize and fairly evaluate visualization code generation. It enforces the use of the same dataset for two chart types to ensure controlled comparison, and requires consistent styling for professional uniformity. 

\begin{figure*}[t!]
\centering
\begin{tcolorbox}[
    colback=gray!2,
    colframe=blue!70!black,
    colbacktitle=blue!12!white,
    title=\textbf{Plot Transformation Data Synthesis Prompt},
    fonttitle=\bfseries,
    coltitle=blue!50!black,
    enhanced,
    sharp corners,
    boxrule=0.5pt,
    left=6pt,
    right=6pt,
    top=5pt,
    bottom=5pt,
    titlerule=0mm,
    width=0.96\textwidth, 
    title style={left color=blue!8!white, right color=blue!5!white}
]
\textbf{System Prompt:}
You are a senior data visualization engineer.
You must synthesize ONE practical significance dataset that works for BOTH specified chart types,
and then output EXACTLY TWO matplotlib code blocks (Code A and Code B), one per chart type.

HARD requirements:\\
- Use matplotlib; numpy is allowed. Do NOT read files.\\
- Declare the SAME dataset inline (lists/ndarrays) in BOTH code blocks.\\
- The two chart types are provided (chart\_pair). Ensure the dataset is semantically suitable for both.\\
- Keep consistent, professional styling (titles, axes labels, legend if relevant, grid/ticks/colormap).\\
- Include plt.show() at the end of each block.\\
- Output ONLY the two python code fences. No extra explanations.\\
Return strictly:

\verb|```python|\\
\verb|# Code A: <first chart type> | \\
\verb|...|\\
\verb| plt.show()|\\
\verb|```|

\verb|```python|\\
\verb|# Code B: <first chart type> | \\
\verb|...|\\
\verb| plt.show()|\\
\verb|```|

\textbf{User Prompt:}\\
chart\_pair:\\
- Chart A: \{chart\_a\} \\
- Chart B: \{chart\_b\}\\

Task:\\
1) Synthesize a meaningful dataset that can be visualized in BOTH chart types above.\\
2) Write two matplotlib code blocks, one for each chart type.\\
3) BOTH blocks must embed and reuse the EXACT SAME dataset (copy-paste identical arrays/values).\\
4) The amount of information in each chart should be as large as possible.\\
5) NO file I/O, NO seaborn, NO CSV reading.\\
6) Keep layout tidy (titles/labels/legend if applicable; suitable colormap for 2D grids; proper axis ticks).\\
7) Output ONLY two code blocks wrapped with\\ \verb|```python|\\
and \verb|```|
\end{tcolorbox}

\caption{Prompt for Plot2Plot Transmation Systhesis Data.}
\label{fig:sytheic}
\end{figure*}

\section{Plot Type Transformation Pairs}
\label{sec:generated_paris}


We manually verified, with the assistance of human experts, the convertible pairs among the 29 categories. Using GPT-5, we then generated approximately approximately 300 data samples, from which human experts further filtered and finalized 27 valid entries. The list of convertible types is presented as follows:
\begin{tcolorbox}[
    colback=gray!2,
    colframe=blue!70!black,
    colbacktitle=blue!12!white,
    title=\textbf{Plot Type Transformation Pairs},
    fonttitle=\bfseries,
    coltitle=blue!50!black,
    enhanced,
    sharp corners,
    boxrule=0.5pt,
    left=6pt,
    right=6pt,
    top=5pt,
    bottom=5pt,
    titlerule=0mm,
    width=0.5\textwidth, 
    title style={left color=blue!8!white, right color=blue!5!white}
]
3D Chart $\rightarrow$ Contour Plot\\
3D Chart $\rightarrow$ Heatmap\\
Area Chart $\rightarrow$ Bar Chart\\
Box Plot $\rightarrow$ Density Plot\\
Box Plot $\rightarrow$ Error Bar Chart\\
Box Plot $\rightarrow$ Violin Plot\\
Bubble Chart $\rightarrow$ 3D Chart\\
Bubble Chart $\rightarrow$ Bar Chart \\
Candlestick Chart $\rightarrow$ Box Plot\\
Candlestick Chart $\rightarrow$ Heatmap\\
Contour Plot $\rightarrow$ 3D Chart \\
Density Plot $\rightarrow$ Error Bar Chart \\
Donut Chart $\rightarrow$ Bar Chart \\
Donut Chart $\rightarrow$ Funnel Chart \\
Funnel Chart $\rightarrow$ Bar Chart \\
Heatmap $\rightarrow$ Contour Plot \\
Histogram $\rightarrow$ Box Plot\\
Line Chart $\rightarrow$ Area Chart\\
Line Chart $\rightarrow$ Bar Chart\\
Line Chart $\rightarrow$ Nightingale Rose Chart\\
Line Chart $\rightarrow$ Radar Chart\\
Pie Chart $\rightarrow$ Bar Chart\\
Radar Chart $\rightarrow$ Bar Chart\\
Scatter Plot $\rightarrow$ 3D Chart\\
Scatter Plot $\rightarrow$ Bubble Chart\\
Violin Plot $\rightarrow$ Density Plot\\
Violin Plot $\rightarrow$ Error Bar Chart
\end{tcolorbox}

\section{Detailed Category Taxonomy and Overview of Chart Type}
\label{sec:detaild category}
\begin{figure*}
    \centering
    \includegraphics[width=0.8\linewidth]{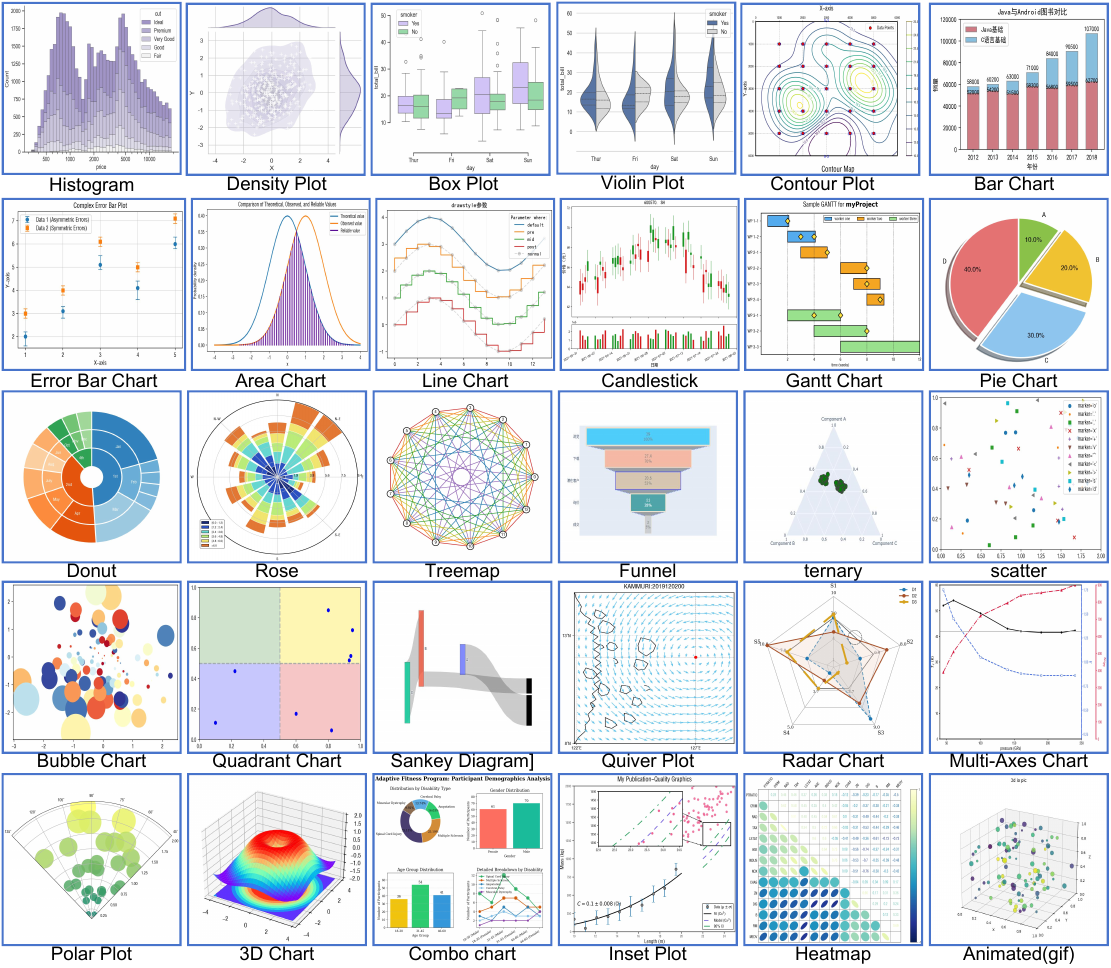}
    \caption{Overview of different types of plots}
    \label{fig:details}
\end{figure*}

An overview of various plot types is shown in Figure \ref{fig:details}. The distribution across the nine categories is summarized as follows: Enhancement (204 instances) dominates the dataset, driven by its large share of 3D charts (99) and combo charts (55). Composition (118) follows, including treemaps, donut charts, and rose charts that emphasize proportion-oriented structures. Distribution (102), Comparison (66), and Trend (63) together represent canonical statistical charts such as histograms, bar charts, and line charts. The remaining categories, Correlation (50), Dimension (49), Flow (22), and Animation (36), contribute additional structural diversity, from scatter plots and radar charts to Sankey diagrams and animated visualizations. This balanced yet skewed coverage ensures that the benchmark captures both widely used and structurally challenging visualization types.

\section{Prompt Template for Verifying the Complexity of the Plot}
\label{sec:evalution_complexity_prompt}
The “Plot Complexity Evaluation Prompt” (See Figure \ref{fig:evalution_complexity_prompt}) is designed to establish a standardized, objective way to assess the visual complexity of plots generated by Python visualization libraries. Its core intention is to shift evaluation from code-level features to what is visually perceived in the image itself, ensuring that models can judge complexity based on appearance rather than implementation details.

By defining three key dimensions, visual density, structural complexity, and styling complexity, and restricting responses to one of three levels (“simple,” “intermediate,” or “complex”), the prompt enables consistent, machine-readable complexity labeling.

\section{Details about Different Python Libraries}
\label{sec:details_about_librabies}
This multi-library composition introduces significant variation in syntax, API abstraction, and rendering behavior: while Matplotlib and Seaborn adopt a procedural and object-oriented approach for static visualization, Plotly and Plotnine employ declarative and event-driven paradigms that require fundamentally different reasoning strategies. 
Additionally, library-specific stylistic conventions, such as default color palettes, annotation grammars, and layout hierarchies, further amplify the diversity of the benchmark, compelling models to generalize beyond superficial syntax to capture underlying visualization intent. 
Such cross-library heterogeneity makes PlotGen-Bench uniquely suited for assessing whether visualization generation models can robustly transfer knowledge across distinct plotting ecosystems.


\begin{figure*}[]
\centering
\begin{tcolorbox}[
    colback=gray!2,
    colframe=blue!70!black,
    colbacktitle=blue!12!white,
    title=\textbf{Plot Complexity Evaluation Prompt},
    fonttitle=\bfseries,
    coltitle=blue!50!black,
    enhanced,
    sharp corners,
    boxrule=0.5pt,
    left=6pt,
    right=6pt,
    top=5pt,
    bottom=5pt,
    titlerule=0mm,
    width=0.96\textwidth, 
    title style={left color=blue!8!white, right color=blue!5!white}
]

 You are an expert in data visualization and chart analysis.
Your task is to evaluate the visual complexity of a plot image generated by Python plotting tools (e.g., Matplotlib, Seaborn, Plotly).
Focus on what is visible in the image — not the underlying code.

Consider the following dimensions:\\
1. Visual density — number of graphical elements, text labels, colors, and annotations.\\
2. Structural complexity — number of subplots, layout composition, or multi-chart arrangement.\\\
3. Styling complexity — color mapping, grid or axis customization, legends, and decorative effects.

There are exactly three valid complexity categories:\\
- "simple": A single, clean plot with minimal visual elements.
- "intermediate": Moderate complexity — multiple elements, customized styles, or subplots.\\
- "complex": High visual density, multiple plot types combined, annotations, or intricate layouts.

Output Rules:
- Respond strictly in JSON — no markdown, no natural language text outside JSON.\\
- The JSON must follow this schema exactly:
\verb|```python|\\
\{
  "complexity": "simple | intermediate | complex",\\
  "reason": "A concise English explanation (max 30 words)"\\
\}\\
\verb|```|
\end{tcolorbox}

\caption{Prompt for evaluating visual complexity}
\label{fig:evalution_complexity_prompt}
\end{figure*}

\section{Instruction examples of different tasks}
\label{sec:instruction_exmaples_differnet_tasks}
\begin{figure*}
    \centering
    \includegraphics[width=0.85\linewidth]{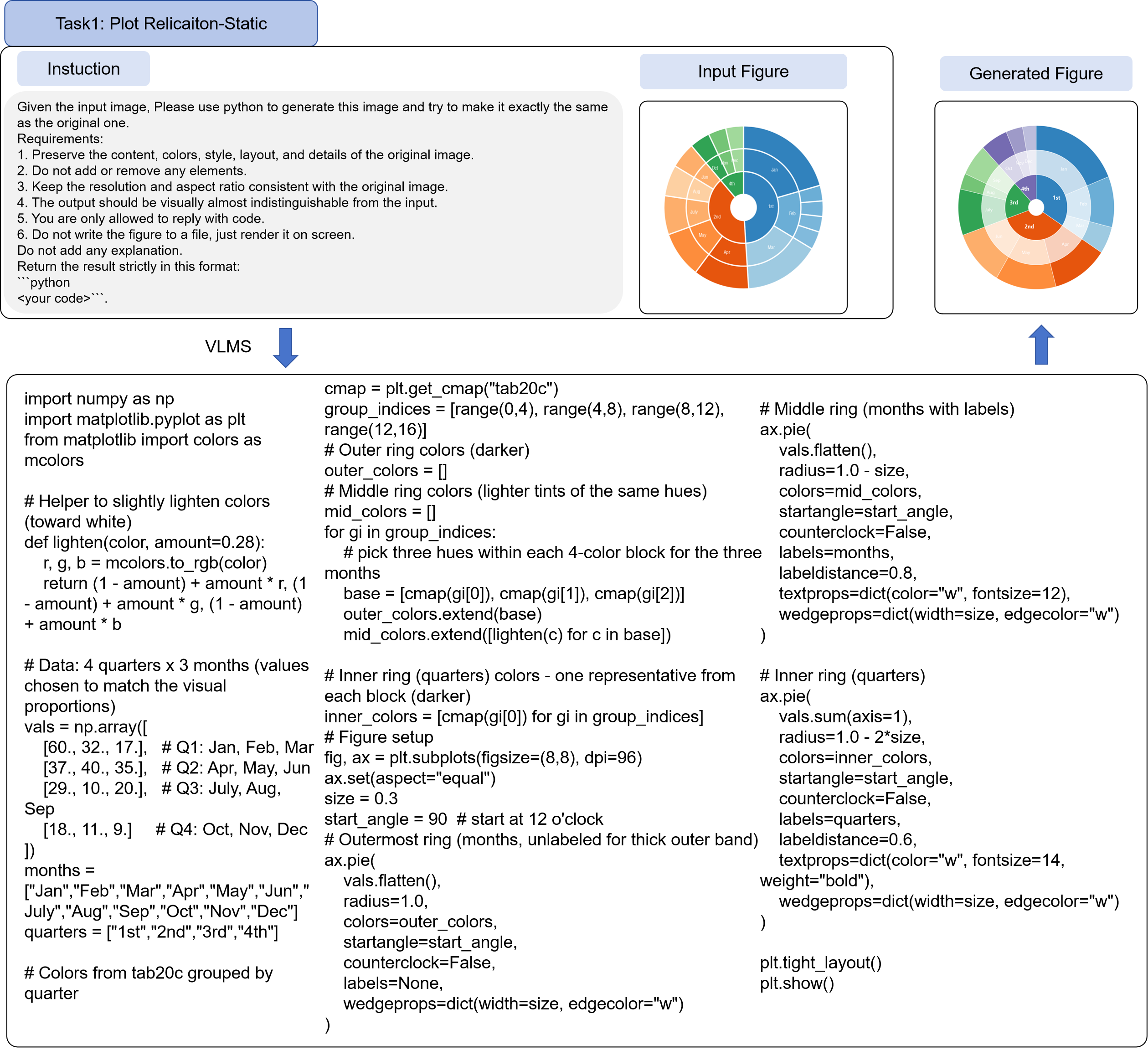}
    \caption{Instruct example of Plot Replication (Static) }
    \label{fig:instruction-1}
\end{figure*}

\begin{figure*}
    \centering
    \includegraphics[width=0.82\linewidth]{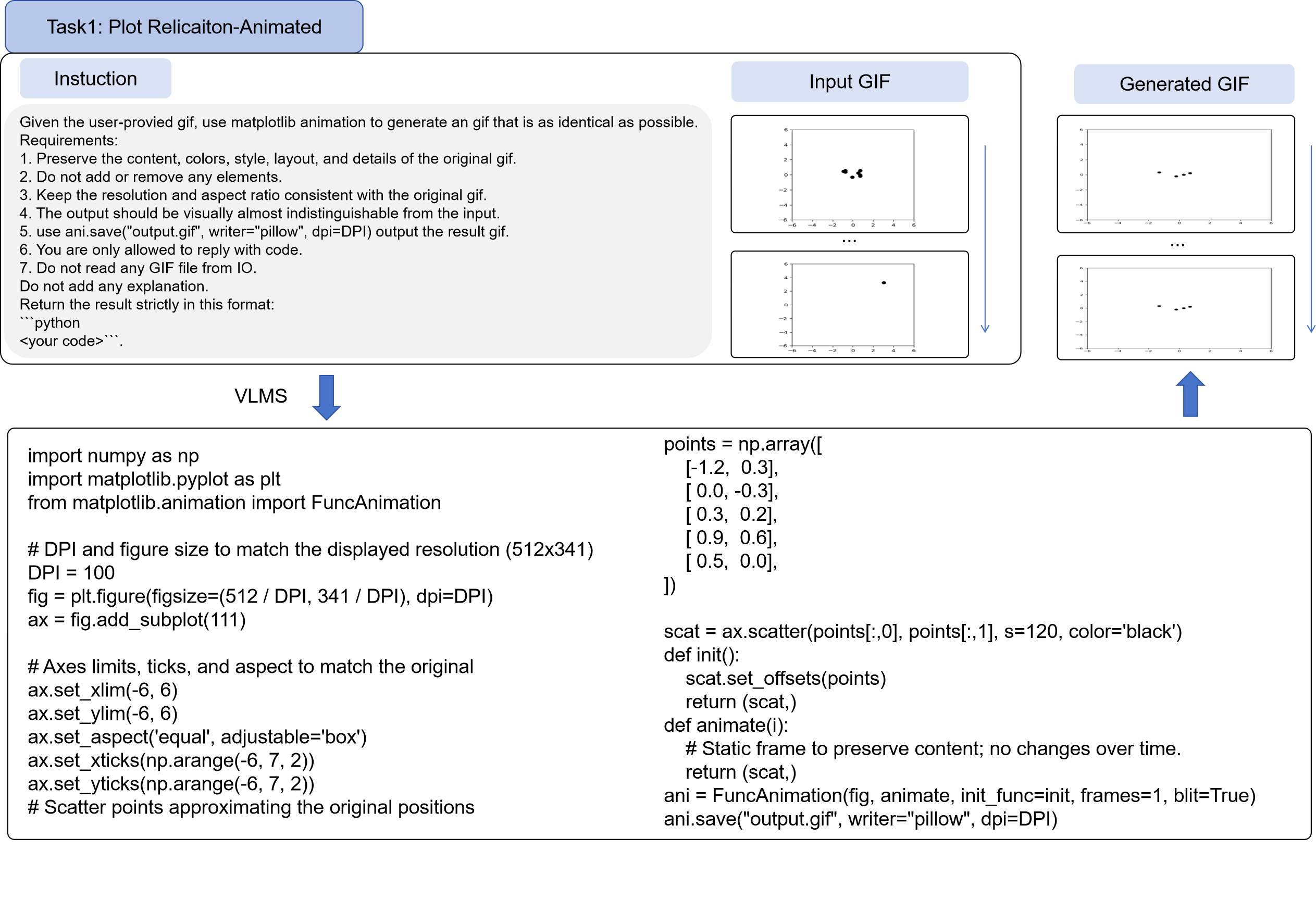}
    \caption{Instruct example of Plot Replication (Animation) }
    \label{fig:instruction-2}
\end{figure*}

\begin{figure*}
    \centering
    \includegraphics[width=0.8\linewidth]{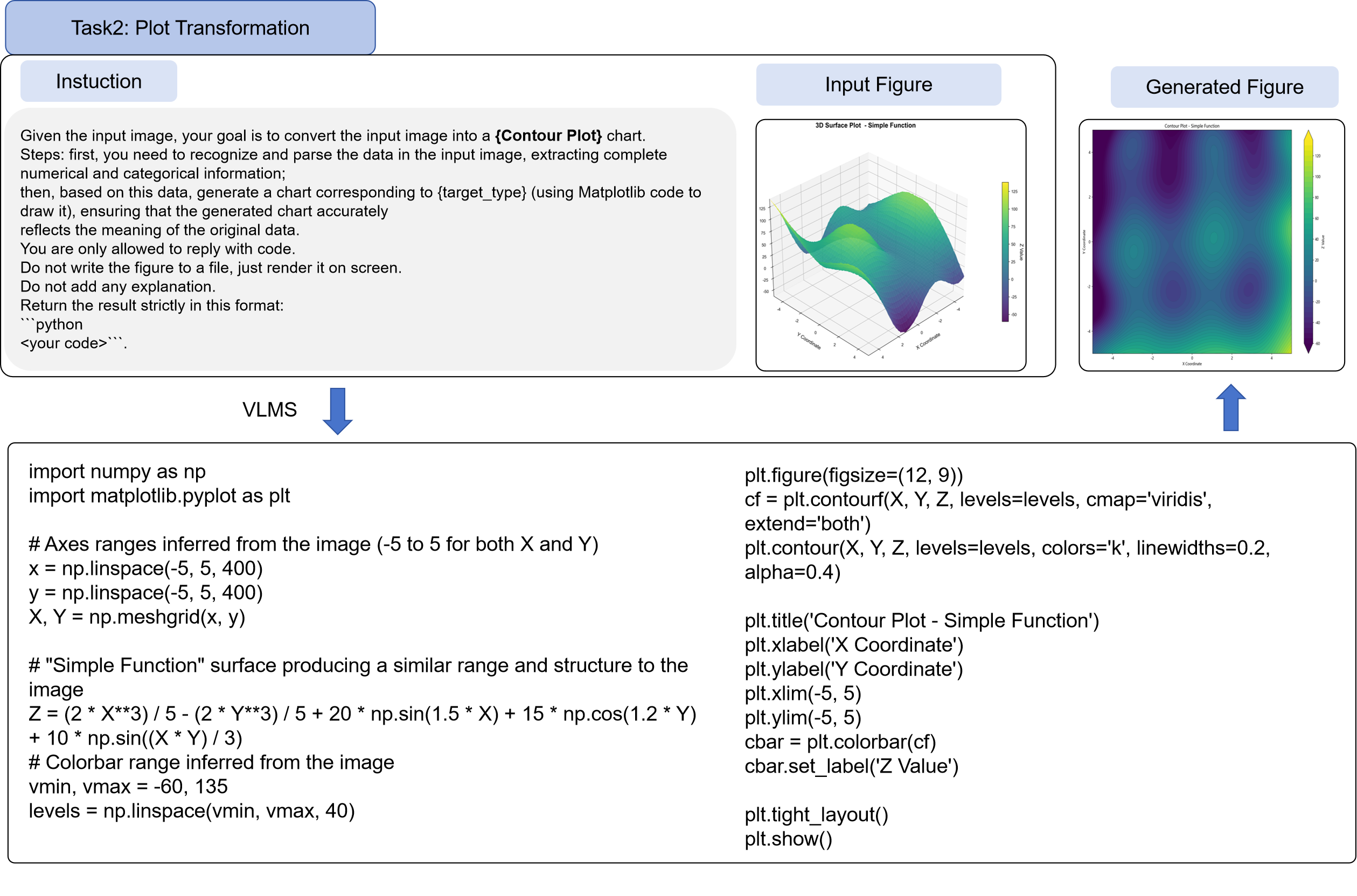}
    \caption{Instruct example of Plot Transformation}
    \label{fig:instruction-3}
\end{figure*}

\begin{figure*}
    \centering
    \includegraphics[width=0.8\linewidth]{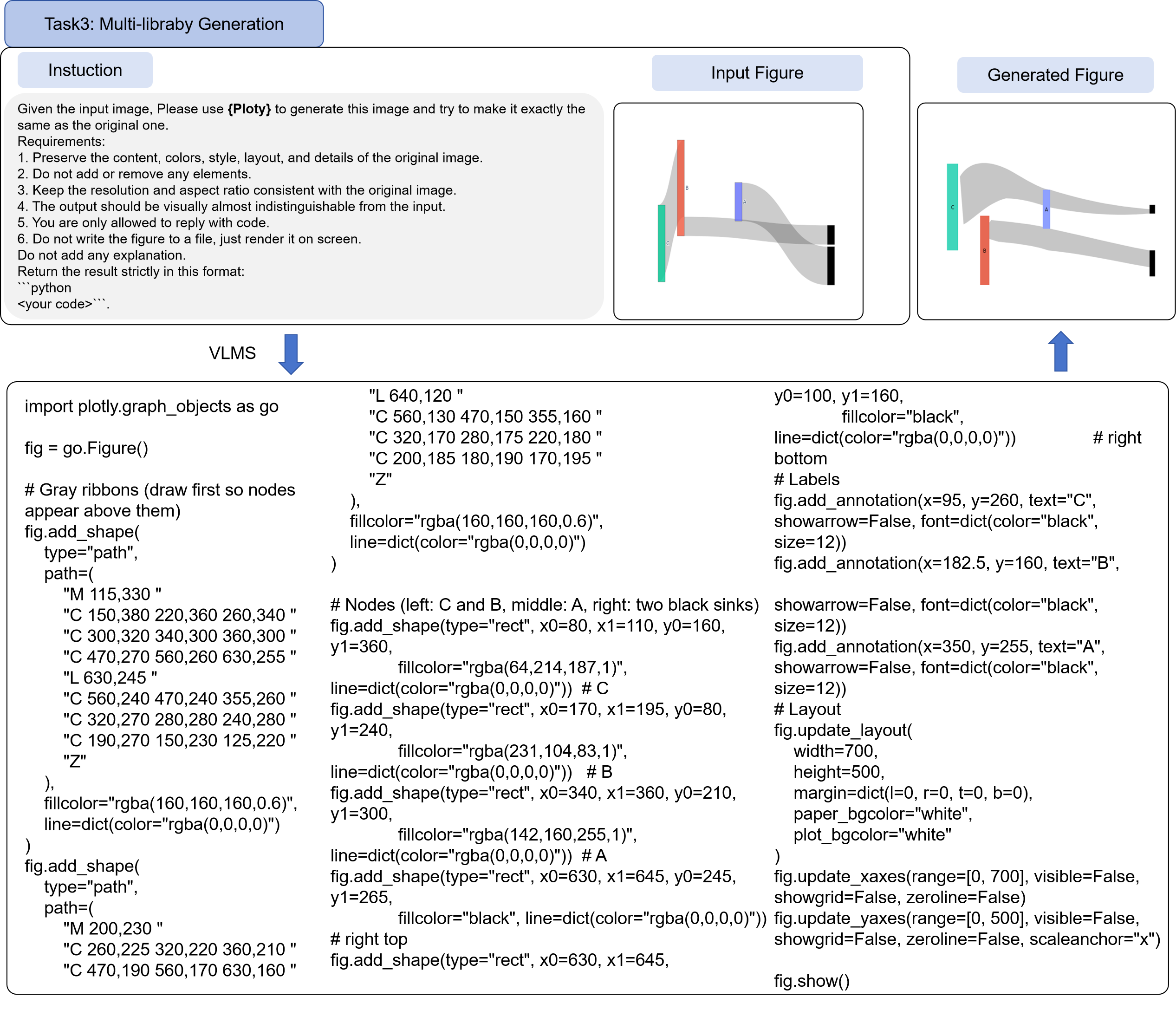}
    \caption{Instruct example of Multi-library Generation}
    \label{fig:instruction-4}
\end{figure*}

We have summarized representative cases of the instruction process for typical tasks, as shown in Figure \ref{fig:instruction-1},\ref{fig:instruction-2},\ref{fig:instruction-3} and \ref{fig:instruction-4}, including how to generate static plots, animated plots, type-transformation plots, and plots using specific visualization libraries.

\section{GPT4O Evaluation Case Study}
\label{sec:evalution_case_study_gpt4o}
\begin{figure*}
    \centering
    \includegraphics[width=1\linewidth]{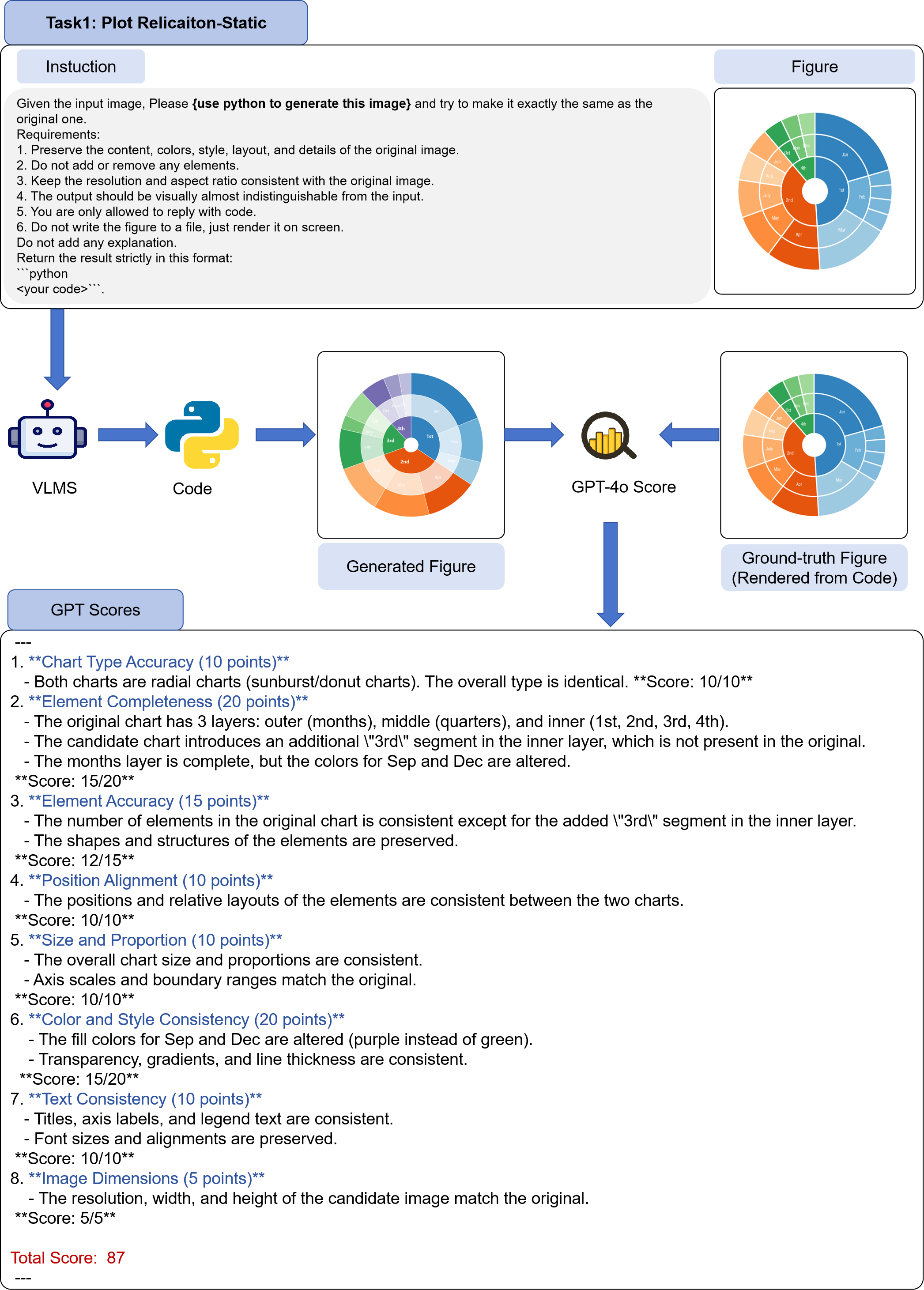}
    \caption{Example of GPT4O's scoring results of static plot without specific library constraints }
    \label{fig:evaluation1}
\end{figure*}

\begin{figure*}
    \centering
    \includegraphics[width=\linewidth]{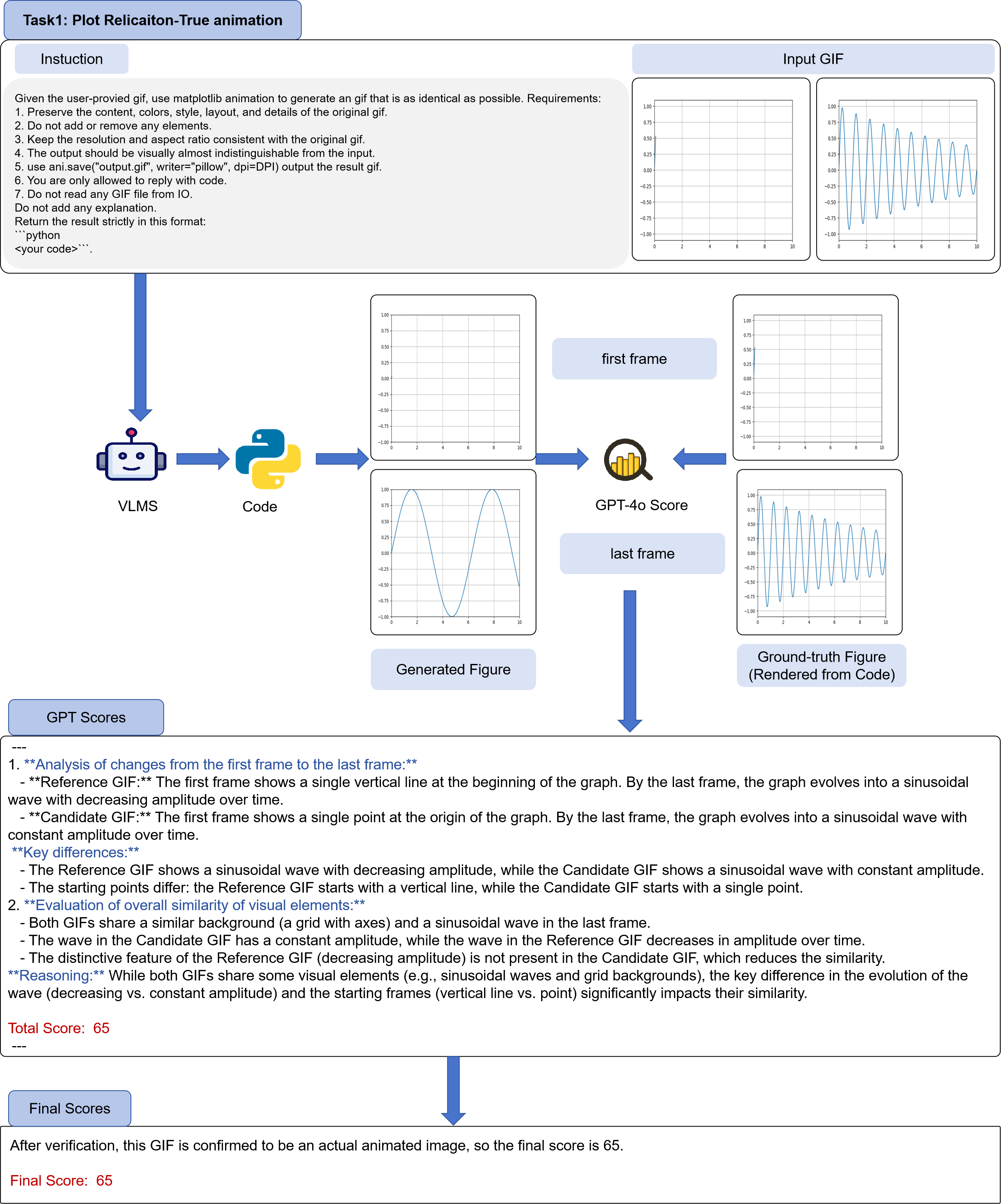}
    \caption{Example of GPT4O's scoring results of pseudo animation plot without specific library constraints }
    \label{fig:evaluation2}
\end{figure*}

\begin{figure*}
    \centering
    \includegraphics[width=\linewidth]{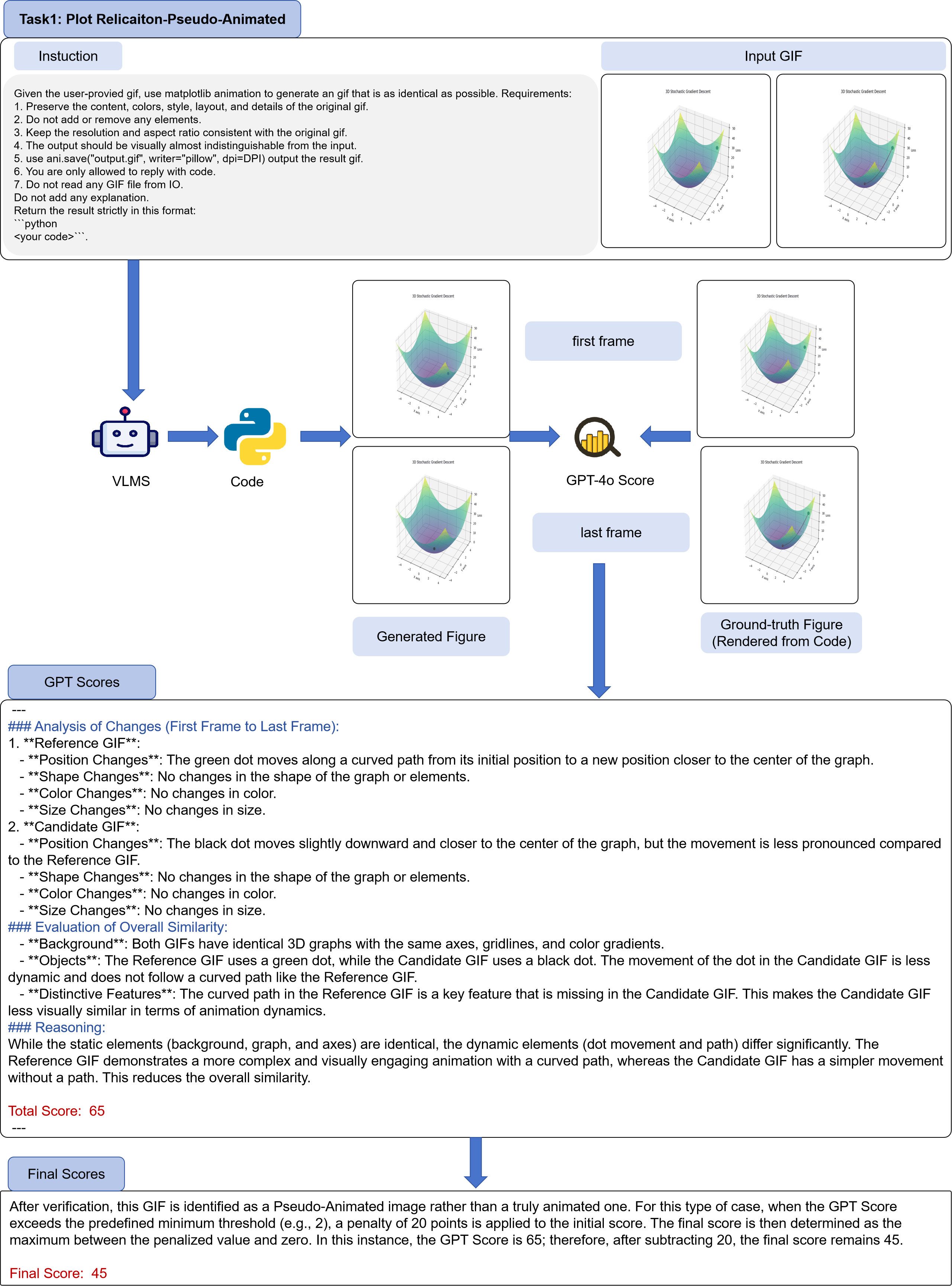}
    \caption{Example of GPT4O's scoring results of true animation plot without specific library constraints }
    \label{fig:evaluation3}
\end{figure*}

\begin{figure*}
    \centering
    \includegraphics[width=\linewidth]{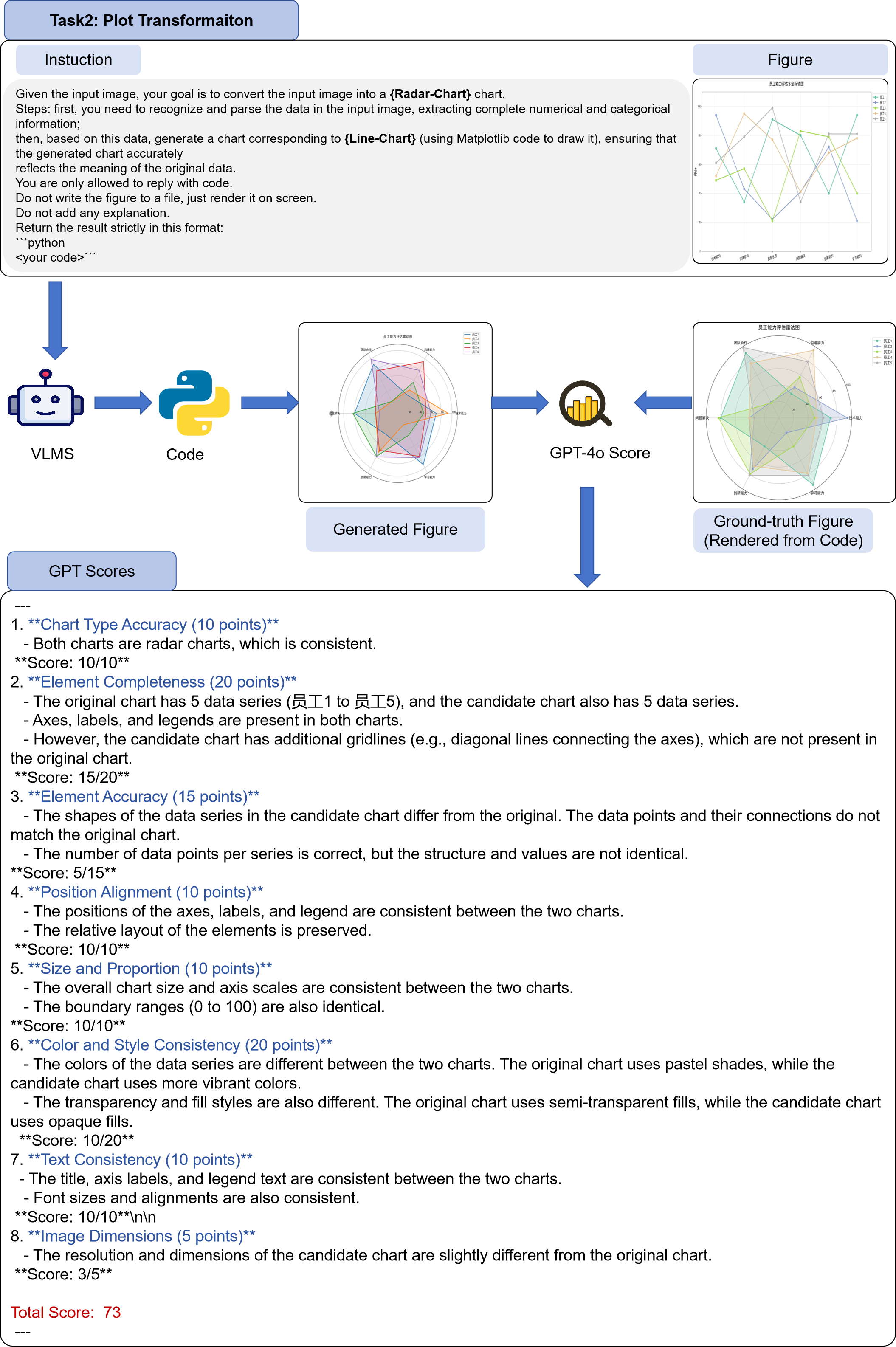}
    \caption{Example of GPT4O's scoring results of plot transformation }
    \label{fig:evaluation4}
\end{figure*}

\begin{figure*}
    \centering
    \includegraphics[width=\linewidth]{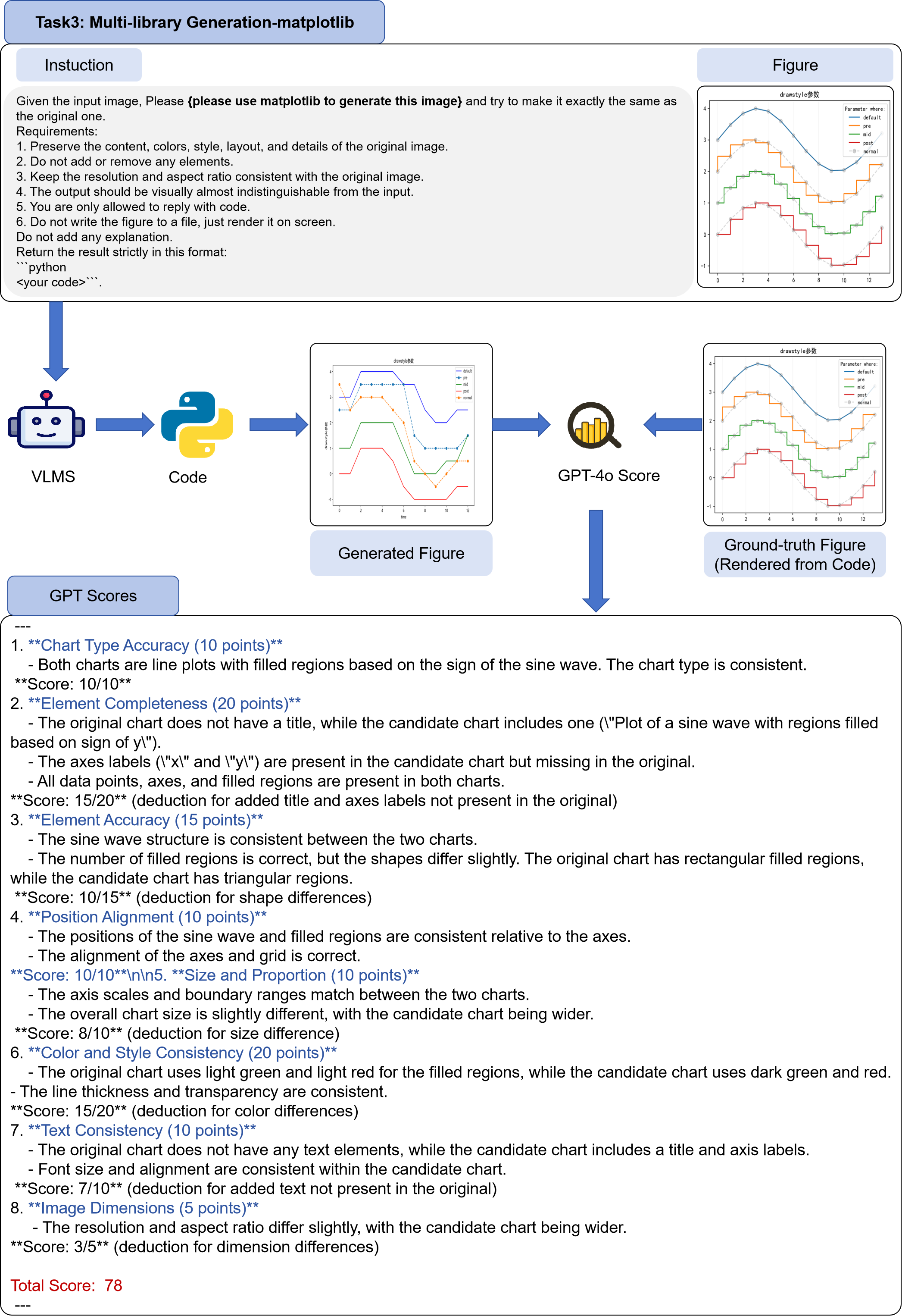}
    \caption{Example of GPT4O's scoring results of static plot with specific library constraints }
    \label{fig:evaluation5}
\end{figure*}

We have aslo summarized representative cases of the instruction process for typical tasks, as shown in Figure \ref{fig:evaluation1},\ref{fig:evaluation2},\ref{fig:evaluation3},\ref{fig:evaluation4} and \ref{fig:evaluation5}, including how to generate static plots, animated plots, type-transformation plots, and plots using specific visualization libraries.


\begin{figure*}[]
\centering
\begin{tcolorbox}[
    colback=gray!2,
    colframe=blue!70!black,
    colbacktitle=blue!12!white,
    title=\textbf{Plot Similarity Evaluation Prompt},
    fonttitle=\bfseries,
    coltitle=blue!50!black,
    enhanced,
    sharp corners,
    boxrule=0.5pt,
    left=6pt,
    right=6pt,
    top=5pt,
    bottom=5pt,
    titlerule=0mm,
    width=0.96\textwidth, 
    title style={left color=blue!8!white, right color=blue!5!white}
]

 You are an expert in verifying chart data consistency. Your task is to strictly compare two images (Original vs Rendered (AI generated)) and, based on the following evaluation criteria, provide a score for each item, note any differences (if applicable), and finally give the total score (out of 100) along with the overall grade.\\
Evaluation Criteria:\\
1.Chart Type Accuracy (10 points): Determine whether the overall type of the rendered chart is identical to the original.\\
2.Element Completeness (20 points): Check if all elements from the original (data points, axes, legends, labels, etc.) are fully preserved in the rendered chart, with no omissions or additions.\\
3.Element Accuracy (15 points): Verify whether the shape, number, and structure of each element are identical to the original.\\
4.Position Alignment (10 points): Confirm whether the positions and relative layouts of all elements are consistent with the original.\\
5.Size and Proportion (10 points): Ensure that the overall chart size, axis scales, and boundary ranges match the original.\\
6.Color and Style Consistency (20 points): Check whether fill colors, line colors, background colors, line thickness, transparency, gradients, etc. are consistent with the original.\\
7.Text Consistency (10 points): Verify whether titles, axis labels, legend text, font sizes, font types, and alignments are consistent with the original.\\
8.Image Dimensions (5 points): Confirm whether the resolution, width, and height of the image match the original.\\
\#\#\#Evaluation\\
After providing your explanation, you must rate the response on a scale of 1 to 100 by strictly following this format:
"Rating: [[85]]".\\
Do not output the score in any other form.
\end{tcolorbox}
\caption{Prompt for evaluating plot similarity}
\label{fig:plot-evalution-similairy-prompt}
\end{figure*}

\begin{figure*}[]
\centering
\begin{tcolorbox}[
    colback=gray!2,
    colframe=blue!70!black,
    colbacktitle=blue!12!white,
    title=\textbf{GIF(Animation) Evaluation Similarity Prompt},
    fonttitle=\bfseries,
    coltitle=blue!50!black,
    enhanced,
    sharp corners,
    boxrule=0.5pt,
    left=6pt,
    right=6pt,
    top=5pt,
    bottom=5pt,
    titlerule=0mm,
    width=0.96\textwidth, 
    title style={left color=blue!8!white, right color=blue!5!white}
]

 You have two animated GIFs, each represented by their first and last frames (four images in total): \\
Reference GIF first frame, Reference GIF last frame, Candidate GIF first frame, Candidate GIF last frame. \\
Please follow this evaluation procedure strictly: \\
1. First, analyze the changes from the first frame to the last frame, considering:\\
   - Position changes of objects/elements\\
   - Shape changes\\
   - Color changes\\
   - Size changes\\
2. Then, evaluate the overall similarity of visual elements between the two frames, including background, objects, and any distinctive features.
Provide a brief explanation of your reasoning. \\
After providing your explanation, you must rate the response on a scale of 1 to 100 by strictly following this format:\\
"Rating: [[85]]".\\
Do not output the score in any other form.
\end{tcolorbox}
\caption{Prompt for evaluating GIF pairs' similarity}
\label{fig:gif-similarity-prompt}
\end{figure*}
\section{Prompt Template for Evaluation}
\label{sec:evaluation_prompt}
The two prompts shown in the Figure \ref{fig:plot-evalution-similairy-prompt} and Figure \ref{fig:gif-similarity-prompt} serve as standardized evaluation guidelines designed to ensure both visual and semantic fidelity between original and generated visualizations. Their overarching goal is to provide a rigorous, human-interpretable framework for assessing how accurately an AI model can reproduce or animate plots compared with the original reference.

The Plot Similarity Evaluation Prompt focuses on static visualization comparison. Its purpose is to evaluate whether a rendered chart faithfully reproduces the structure, layout, and stylistic details of the original image. The evaluation criteria cover eight dimensions, including chart type accuracy, element completeness, element accuracy, position alignment, size and proportion, color and style consistency, text consistency, and image dimensions. Together, these metrics provide a fine-grained and quantitative assessment of visual correspondence, emphasizing both the semantic and aesthetic aspects of chart replication. This prompt is designed to simulate the judgment of a professional data-visualization expert, ensuring that the generated chart is not only executable but visually equivalent to the reference.

The GIF (Animation) Evaluation Similarity Prompt, in contrast, is tailored for assessing dynamic visualizations. It evaluates the degree of similarity between two animated GIFs—specifically by comparing their first and last frames—to capture how faithfully the candidate animation reproduces the temporal and structural transitions of the reference. The procedure examines position, shape, color, and size changes across frames, as well as the overall visual coherence of motion and transformation. By doing so, it shifts the evaluation focus from static fidelity to dynamic consistency, measuring how accurately the animation conveys continuous visual evolution.

\section{The Details of Animation Evaluation}
\label{sec:animation_evalution}
We have already mentioned in the main text that the final score is computed from two components: one based on GPT-4o evaluation, denoted as $GS$, and the other based on pHash to determine whether the GIF is a pseudo-animation. These two components are combined as follows: if the GIF is identified as a pseudo-animation, we subtract 20 points from $GS$. To avoid negative scores for validly generated GIFs, we further apply a lower bound by taking $max(2, GS - 20)$. This ensures that even penalized outputs retain a minimal positive score, thereby encouraging the model to generate correct code. If the GIF is classified as a genuine animation, no penalty is applied, and the score remains $GS$.

\section{Code-related Error Definition}
\label{sec:code-related-error-definition}
This appendix provides concise definitions for the five categories of exceptions observed in code execution, as referenced in Figure~\ref{fig:code-related-errors}.

\noindent \textbf{TypeError} \\
A TypeError occurs when an operation or function is applied to an object of an inappropriate type. This includes mismatches between actual argument types and function signatures (e.g., passing a list where a float is expected), or conflicts between positional and keyword arguments at the type level.
\begin{lstlisting}[language=Python, caption=An exemplary Python code snippet illustrating a TypeError., label=lst:text]
import matplotlib.pyplot as plt
from matplotlib.patches import Patch

# Figure size to match the provided image resolution (approx 886x768 px at 100 DPI)
fig, ax = plt.subplots(figsize=(8.86, 7.68), dpi=100)

# Data
sizes_outer = [10, 20, 30, 40]        # Outer ring percentages
sizes_inner = [4, 6, 8, 10]           # Inner ring values -> 14.3%, 21.4%, 28.6%, 35.7%
labels = ['A', 'B', 'C', 'D']
colors = ['blue', 'green', 'red', 'cyan']

# Outer donut
wedges_outer, texts_outer, autotexts_outer = ax.pie(
    sizes_outer,
    radius=1.0,
    colors=colors,
    startangle=90,
    counterclock=True,
    autopct='%.1f%%',
    pctdistance=0.85,
    wedgeprops=dict(width=0.35, edgecolor='black', linewidth=1),
    textprops=dict(color='black', fontsize=12)
)

# Inner donut
wedges_inner, texts_inner, autotexts_inner = ax.pie(
    sizes_inner,
    radius=0.70,
    colors=colors,
    startangle=90,
    counterclock=True,
    autopct='%.1f%%',
    pctdistance=0.75,
    wedgeprops=dict(width=0.35, edgecolor='black', linewidth=1),
    textprops=dict(color='black', fontsize=12)
)

# Center hole (white)
centre_circle = plt.Circle((0, 0), 0.35, color='white')
ax.add_artist(centre_circle)

# Legend on the right
legend_elements = [Patch(facecolor=c, edgecolor='black', label=l) for c, l in zip(colors, labels)]
leg = ax.legend(legend_elements, loc='center left', bbox_to_anchor=(0.93, 0.5), frameon=True)
leg.get_frame().set_edgecolor('black')
leg.get_frame().set_linewidth(1)

ax.axis('equal')
plt.subplots_adjust(left=0.08, right=0.88, top=0.95, bottom=0.08)

plt.show()
\end{lstlisting}

\noindent
The error arises because the \texttt{ax.legend()} method in 
\texttt{Matplotlib} is defined as 
\texttt{ax.legend(handles, labels, **kwargs)}. 
When only one positional argument is provided, it is interpreted as 
a list of \textit{labels} rather than a list of \textit{handles}. 
In this case, the variable \texttt{legend\_elements} contains a list 
of \texttt{Patch} objects (i.e., instances of \texttt{Artist}), 
rather than strings representing labels. Consequently, 
\texttt{Matplotlib} raises a \texttt{TypeError} due to a type mismatch 
between the expected and actual argument types.

\noindent

To resolve this issue, the parameter name should be explicitly specified 
to avoid semantic ambiguity, as shown below:

\begin{scriptsize}
\begin{verbatim}
leg = ax.legend(handles=legend_elements, loc='center left',
                bbox_to_anchor=(0.93, 0.5), frameon=True)
\end{verbatim}
\end{scriptsize}

\noindent \textbf{ValueError} \\
A ValueError is raised when a function receives an argument of the correct type but with an inappropriate value. This includes out-of-range values, empty data, conflicting parameters, or inconsistent configurations (e.g., bins not being sorted, or the number of colors not matching the number of series).
\begin{lstlisting}[language=Python, caption=An exemplary Python code snippet illustrating a ValueError., label=lst:text]
import numpy as np
import matplotlib.pyplot as plt
from matplotlib.projections.polar import PolarAxes
from matplotlib.projections import register_projection
from matplotlib.path import Path
from matplotlib.spines import Spine
from matplotlib.transforms import Affine2D
from matplotlib import rcParams, font_manager

# Try to use a Chinese-capable font if available
preferred_fonts = [
    "Microsoft YaHei", "SimHei", "Noto Sans CJK SC", "Source Han Sans SC",
    "PingFang SC", "WenQuanYi Zen Hei", "Hiragino Sans GB",
    "Arial Unicode MS", "STHeiti", "Songti SC"
]
available = {f.name for f in font_manager.fontManager.ttflist}
for f in preferred_fonts:
    if f in available:
        rcParams["font.sans-serif"] = [f]
        break
rcParams["axes.unicode_minus"] = False

# Labels (clockwise starting from top)
labels = ['Office', 'Restroom', 'Reception', 'Cafeteria', 'Lounge', 'Meeting Room']
N = len(labels)
theta = radar_factory(N, frame='polygon')

# Data:
# - Filled baseline (regular hexagon at radius 3)
# - Foreground series (blue line) resembles the original shape
baseline = [3, 3, 3, 3, 3, 3]
series =   [5, 2, 3, 2, 5, 3]

fig = plt.figure(figsize=(8.3, 7.68), dpi=100)
ax = plt.subplot(111, projection='radar')
ax.set_rmax(5)
ax.set_rticks([1, 2, 3, 4, 5])
ax.set_yticklabels([])

# Grid styling: polygon rings and radial spokes in light gray
grid_color = '#cfcfcf'
ax.yaxis.grid(True, color=grid_color, linewidth=1)
ax.xaxis.grid(True, color=grid_color, linewidth=1)

# Category labels
ax.set_varlabels(labels, fontsize=18, color='#666666')

# Fill baseline polygon
ax.plot(theta, baseline+baseline[:1] , color='#77c2b8', linewidth=0)
ax.fill(theta, baseline, facecolor='#77c2b8', alpha=0.28)

# Foreground series line
ax.plot(theta, series, color='#5da5da', linewidth=3.0)
ax.scatter(theta, series, s=55, color='#8ab6e6', edgecolor='#5da5da', linewidth=2, zorder=3)

# Make outer frame bold and black
ax.patch.set_edgecolor('black')
ax.patch.set_linewidth(2.0)
ax.set_facecolor('white')

plt.tight_layout()
plt.show()

\end{lstlisting}

\noindent
The \texttt{ValueError} arises because the angle vector $\theta$ and the radial
data vector have inconsistent lengths in the call to \texttt{ax.plot()}. In
radar charts, the polygon must be closed by repeating the first data point at
the end of the sequence, ensuring visual continuity of the shape. However, only
the radial values were extended (\texttt{baseline + baseline[:1]}), while the
corresponding angular vector $\theta$ remained unmodified. This mismatch
between $(N)$ and $(N{+}1)$ elements triggers the dimension error. In essence,
the additional point is required to close the radar polygon, but both the
angular and radial sequences must be extended consistently to maintain equal
lengths.

\noindent \textbf{AttributeError} \\
An AttributeError is raised when an attribute or method is accessed on an object that does not possess it. This includes calling non-existent methods or using incorrect keyword arguments (e.g., passing fontsize= to a Text.set method that does not accept it).
\begin{lstlisting}[language=Python, caption=An exemplary Python code snippet illustrating a AttributeError., label=lst:text]
# Figure size close to the provided image (pixels ~772x768)
fig = plt.figure(figsize=(7.72, 7.68), dpi=100, facecolor="white")

# Create directions concentrated around 16 sectors (every 22.5 degrees)
n_samples = 2400
sector_centers = np.arange(0, 360, 22.5)
choices = rng.integers(0, len(sector_centers), size=n_samples)
# Small angular noise to keep narrow petals
wd = (sector_centers[choices] + rng.normal(0, 7.5, size=n_samples)) % 360

# Wind speeds with most values between 1 and 7, some >9
ws = np.clip(rng.rayleigh(scale=3.2, size=n_samples) + rng.uniform(0, 1, size=n_samples), 0, None)

# Figure size close to the provided image (pixels ~772x768)
fig = plt.figure(figsize=(7.72, 7.68), dpi=100, facecolor="white")

# Create windrose axes and position similar to the original layout
ax = WindroseAxes.from_ax(fig=fig)
ax.set_position([0.06, 0.06, 0.88, 0.88])

# Bins and colormap (black->red->orange->yellow->white)
bins = np.arange(0.0, 10.0, 1.0)
ax.bar(
    wd,
    ws,
    bins=bins,
    nsector=16,
    normed=False,
    opening=0.8,
    edgecolor="white",
    lw=0.4,
    cmap=plt.cm.hot
)

# Compass labels to match the style (with hyphens)
angles = np.arange(0, 360, 45)
labels = ["N", "N-E", "E", "S-E", "S", "S-W", "W", "N-W"]
ax.set_thetagrids(angles, labels=labels, fontsize=12, color="#6e6e6e")

# Radial grid styling and ticks
for gl in ax.yaxis.get_gridlines():
    gl.set_color("#bcbcbc")
    gl.set_linewidth(1.0)
    gl.set_alpha(1.0)

# Force rmax and ticks like in the reference image
ax.set_rmax(176.0)
rticks = np.linspace(35.2, 176.0, 5)
ax.set_rgrids(rticks, angle=25, labels=[f"{t:.1f}" for t in rticks], fontdict={"size":11, "color":"#6e6e6e"})
ax.yaxis.set_major_formatter\
(mpl.ticker.FormatStrFormatter("%.1f"))

# Outer circle styling
for spine in ax.spines.values():
    spine.set_linewidth(1.6)
    spine.set_color("black")

# Legend placement and look
leg = ax.set_legend(
    loc="lower left",
    bbox_to_anchor=(0.0, 0.0),
    fontsize=10,
    frameon=True,
)
\end{lstlisting}

\noindent
The \texttt{AttributeError} arises from passing an unsupported keyword argument to a method that does not accept it. In this case, the \texttt{fontdict} parameter used in \texttt{ax.set\_rgrids()} is incompatible with Matplotlib’s
internal \texttt{Text.set()} method, which does not recognize this argument. This mismatch in accepted parameters leads to an attribute access error during execution.

 \noindent \textbf{KeyError/IndexError} \\
A KeyError (or IndexError) occurs when attempting to access a key in a dictionary (or an index in a sequence) that does not exist. This typically happens when accessing dict or Pandas DataFrame columns with non-existent names.
\begin{lstlisting}[language=Python, caption=An exemplary Python code snippet illustrating a  KeyError., label=lst:text]
import numpy as np
import matplotlib as mpl
import matplotlib.pyplot as plt
from mpl_toolkits.mplot3d import Axes3D  # noqa: F401

# Schwefel function (2D)
def schwefel(X, Y):
    d = 2
    return 418.9829 * d - (X * np.sin(np.sqrt(np.abs(X))) + Y * np.sin(np.sqrt(np.abs(Y))))

# Figure and style to closely match the reference
mpl.rcParams.update({
    "font.family": "serif",
    "axes.edgecolor": "black",
    "axes.linewidth": 1.0,
    "xtick.color": "black",
    "ytick.color": "black",
    "ztick.color": "black",
    "figure.facecolor": "white",
    "axes.facecolor": "white",
})

# Create grid (domain commonly used for Schwefel)
x = np.linspace(-450, 450, 320)
y = np.linspace(-450, 450, 320)
X, Y = np.meshgrid(x, y)
Z = schwefel(X, Y)

# Prepare figure sized similarly to the input image
fig = plt.figure(figsize=(7.68, 8.03), dpi=100)
ax = fig.add_subplot(111, projection='3d')

# Set viewing angle to match perspective
ax.view_init(elev=26, azim=-56)
ax.dist = 9.5

# Plot surface
surf = ax.plot_surface(
    X, Y, Z,
    rstride=3, cstride=3,
    cmap="coolwarm",
    linewidth=0.15,
    antialiased=True,
    alpha=0.95
)
\end{lstlisting}

\noindent
The \texttt{KeyError} is raised because the configuration dictionary
\texttt{mpl.rcParams} does not contain the specified key
\texttt{"ztick.color"}. Matplotlib’s runtime configuration only supports
predefined parameter names, and any attempt to update it with an invalid or
non-existent key results in a \texttt{KeyError}. In this case, the
\texttt{"ztick.color"} entry is not part of the valid \texttt{rcParams} set,
leading to the error during the update operation.

 \noindent\textbf{SyntaxError} \\
A SyntaxError indicates a structural error in the code itself, such as missing parentheses or colons, incorrect indentation, or unclosed strings.
\begin{lstlisting}[language=Python, caption=An exemplary Python code snippet illustrating a  SyntaxError., label=lst:text]
import numpy as np
import matplotlib.pyplot as plt

# Figure setup to match aspect/resolution closely (approx 1031x768 px)
fig, ax = plt.subplots(figsize=(10.31, 7.68), dpi=100)

# Grid data
x_labels = [f"A{i}" for i in range(1, 11)]
y_labels = [f"B{i}" for i in range(1, 11)]
X, Y = np.meshgrid(np.arange(1, 11), np.arange(1, 11))
x = X.flatten()
y = Y.flatten()

# Create intensity values: high on the left, lower on the right, slight row-wise variation
# Keep values in ~[0.1, 0.9] to match colorbar range
row_variation = 0.25*np.sin((Y-1)/9*np.pi) + 0.05*np.cos((Y-1)/9*np.pi*2)
col_gradient = 1 - (X-1)/9
intensity = 0.12 + 0.78 * (0.55*col_gradient + 0.45*(0.5+row_variation))
intensity = np.clip(intensity, 0.1, 0.9).flatten()

# Bubble sizes: emphasize area near (A9,B5), with a couple secondary clusters
S = (
    0.95*np.exp(-(((X-9)**2)/3.5 + ((Y-5)**2)/6.0))) +
    0.55*np.exp(-(((X-2)**2)/7.0 + ((Y-9)**2)/6.0))) +
    0.50*np.exp(-(((X-5)**2)/6.5 + ((Y-10)**2)/3.5))) +
    0.30*np.exp(-(((X-2)**2)/4.5 + ((Y-2)**2)/4.5))) +
    0.20*np.exp(-(((X-4)**2)/7.5 + ((Y-7)**2)/8.0)))
S = 200 + 700 * (S.flatten())
S = np.clip(S, 150, 900)

\end{lstlisting}

\noindent
The \texttt{SyntaxError} occurs due to unmatched parentheses in the expression
defining \texttt{S}. The extra opening brackets break the syntactic structure,
causing Python to detect the error during code parsing rather than execution.

\begin{figure}[htbp]
    \centering
    \begin{subfigure}[b]{0.23\textwidth}
        \centering
        \includegraphics[width=\textwidth]{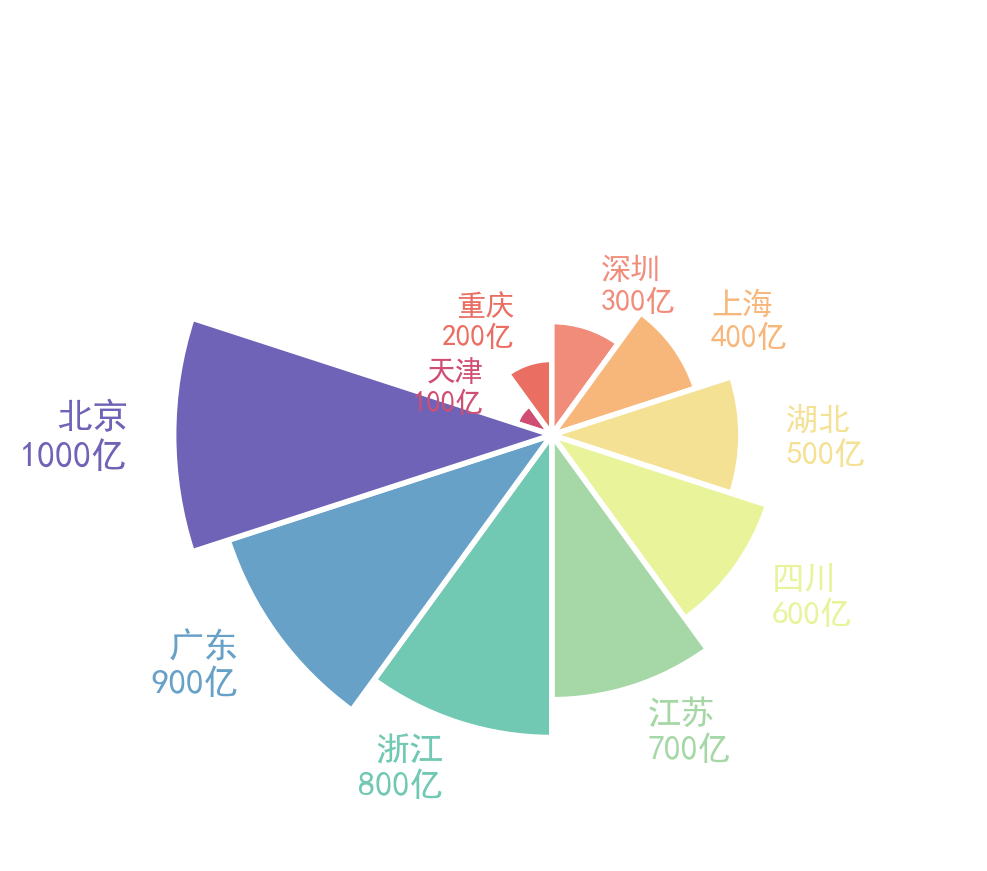}
        \caption{Ground-truth plot}
        \label{fig:code-related-errors}
    \end{subfigure}
    \hfill
    \begin{subfigure}[b]{0.23\textwidth}
        \centering
        \includegraphics[width=\textwidth]{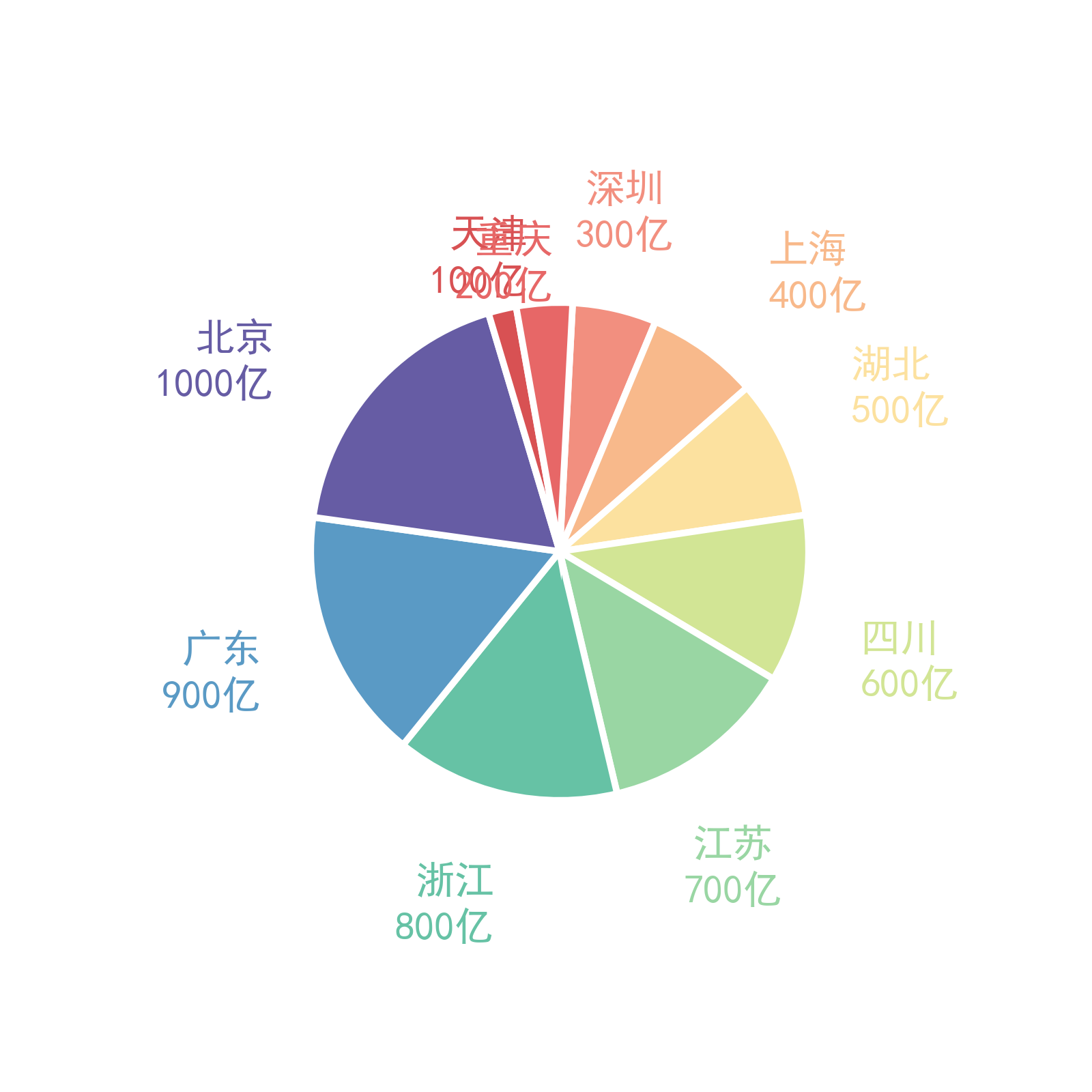}
        \caption{Generated plot}
        \label{fig:render-case4}
    \end{subfigure}
    \caption{Error Case1. In this example, the main error is that the model converted the windrose chart into a pie chart. }
    \label{fig:render-case1}
\end{figure}
\section{Rendering-related Error Definition and Case Study}
\label{sec:rendering-related-error-definition}

This appendix provides concise definitions for the six categories of rendering-related errors observed in generated visual outputs. Specific examples can be found in Figure \ref{fig:render-case1}.

 \noindent \textbf{Chart Type Error} \\
The generated chart type does not match the target requirement (e.g., a line chart is requested but a bar chart is produced).

\begin{figure}[htbp]
    \centering
    \begin{subfigure}[b]{0.23\textwidth}
        \centering
        \includegraphics[width=\textwidth]{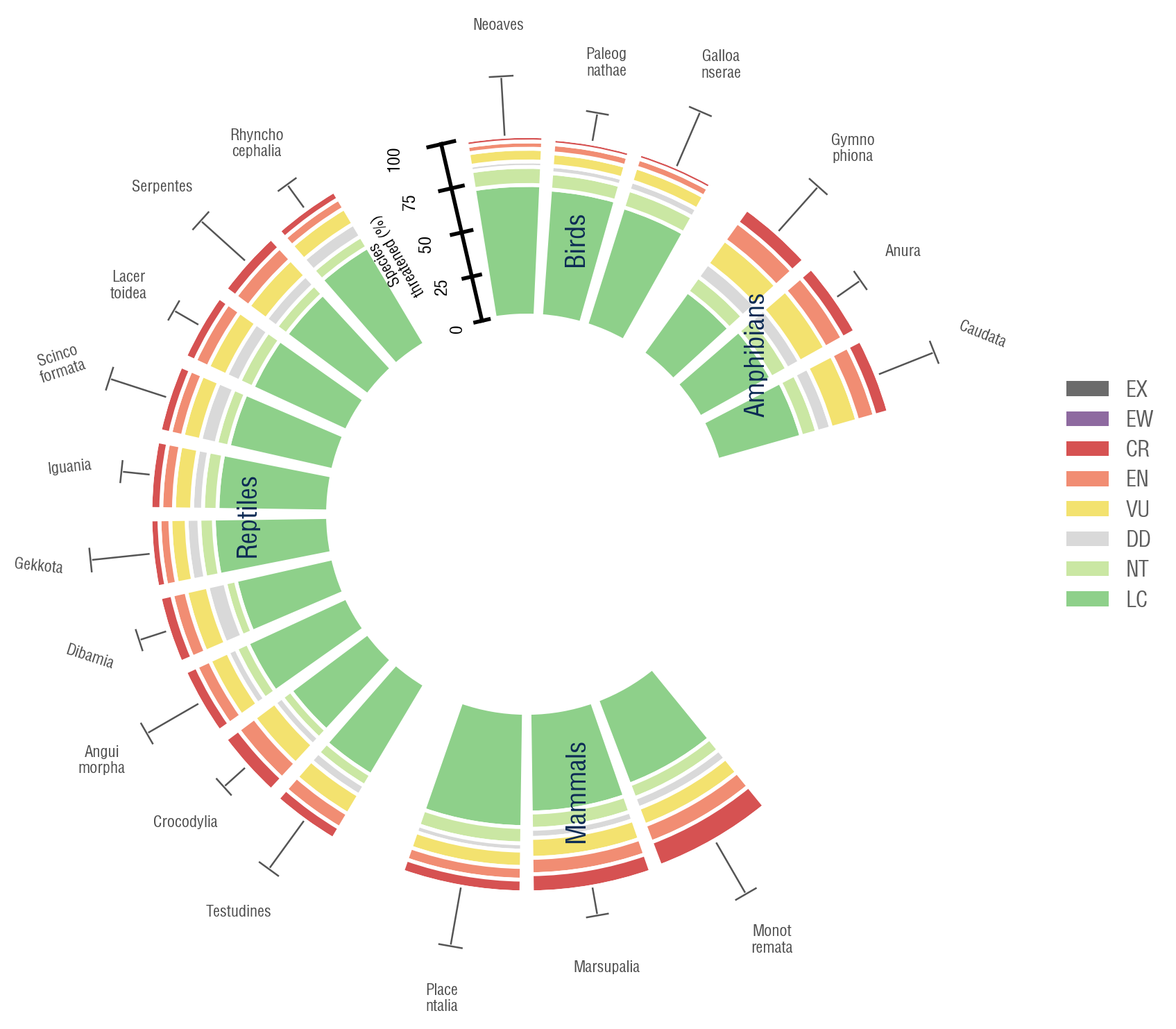}
        \caption{Ground-truth plot}
        \label{fig:code-related-errors}
    \end{subfigure}
    \hfill
    \begin{subfigure}[b]{0.23\textwidth}
        \centering
        \includegraphics[width=\textwidth]{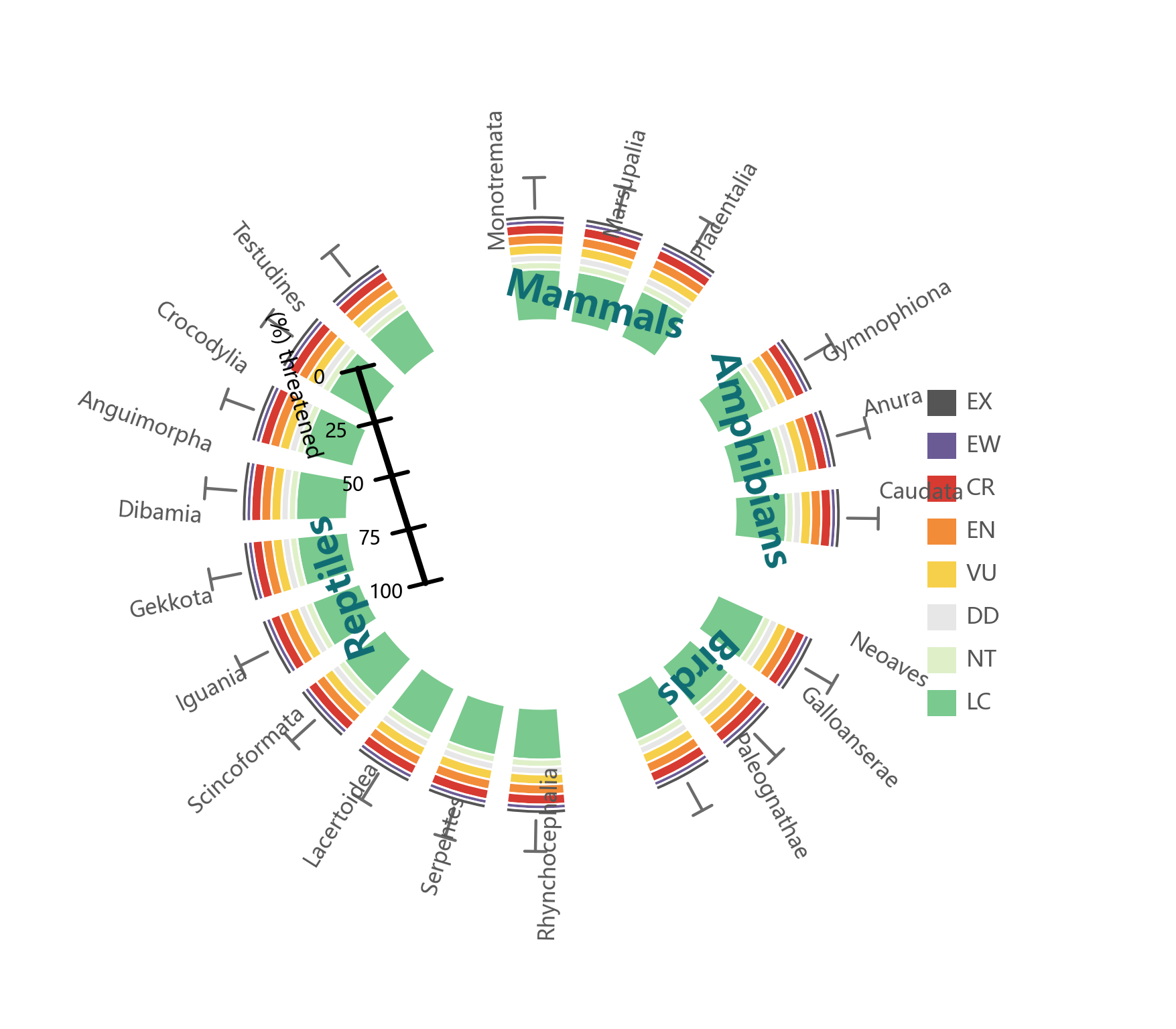}
        \caption{Generated plot}
        \label{fig:}
    \end{subfigure}
    \caption{Rendering Error Case2. In this example, the primary issue lies in Position Alignment, as one key axis and a large number of elements are misaligned. In addition, there are minor color inconsistencies. }
    \label{fig:render-case2}
\end{figure}

 \noindent \textbf{Position Alignment Error} \\
The positions of elements such as axes, legends, or annotations are misaligned or inconsistent with the expected layout. Specific examples can be found in Figure \ref{fig:render-case2}.
\begin{figure}[!htbp]
    \centering
    \begin{subfigure}[b]{0.23\textwidth}
        \centering
        \includegraphics[width=\textwidth]{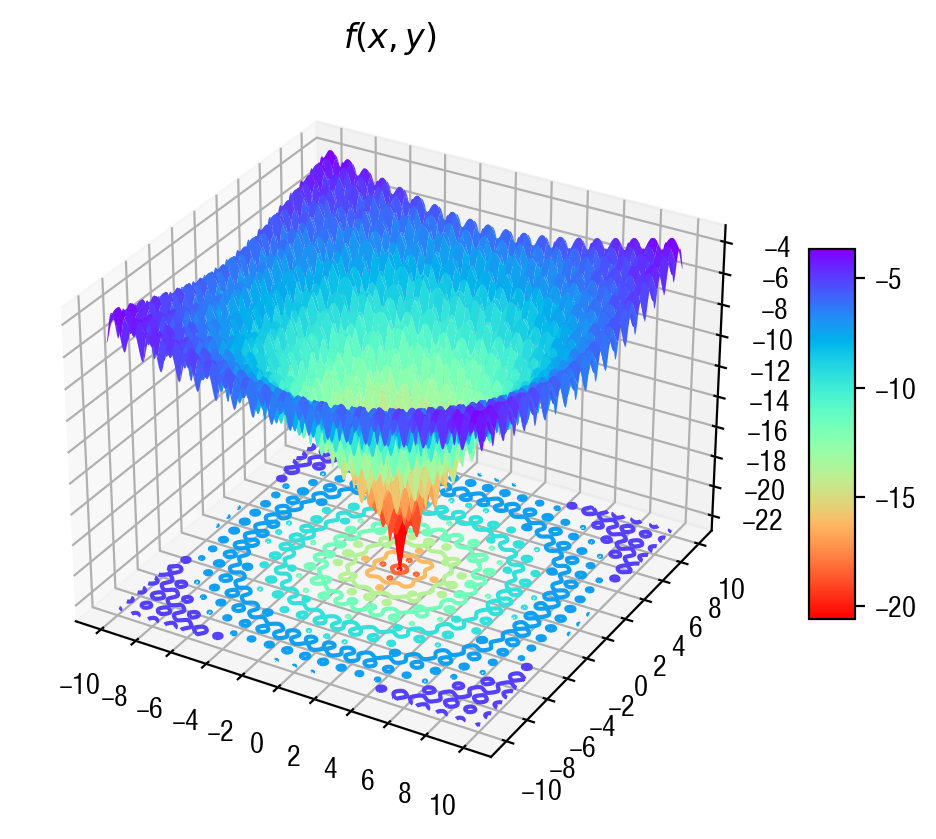}
        \caption{Ground-truth plot}
        \label{fig:code-related-errors}
    \end{subfigure}
    \hfill
    \begin{subfigure}[b]{0.23\textwidth}
        \centering
        \includegraphics[width=\textwidth]{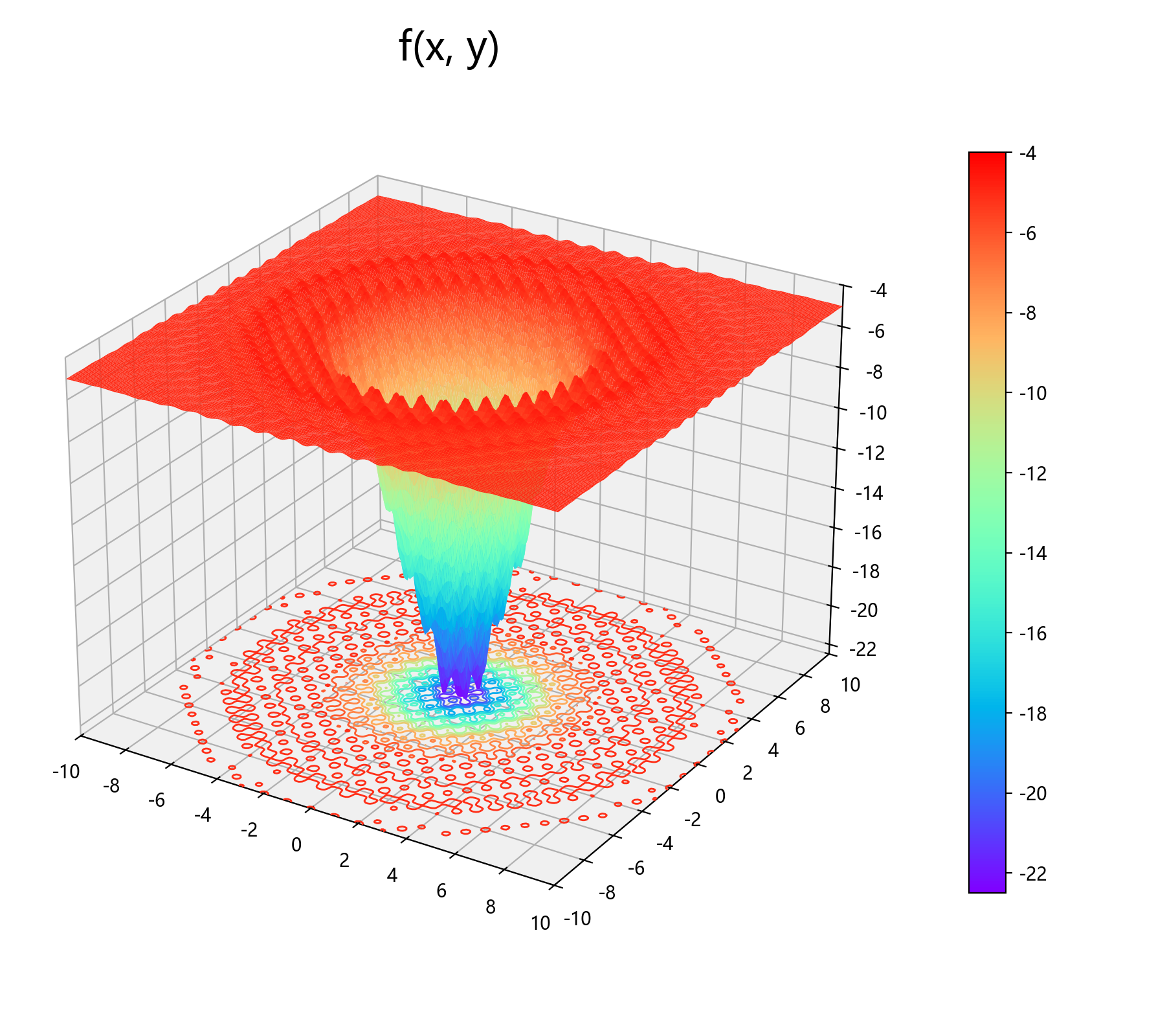}
        \caption{Generated plot}
        \label{fig:}
    \end{subfigure}
    \caption{Error Case3. In this example, the primary error lies in the excessive color deviation, followed by imperfect style alignment. }
    \label{fig:render-case3}
\end{figure}
 \noindent \textbf{Color \& Style Error} \\
The selected colors differ significantly from the original, or the line styles are incorrectly rendered. Specific examples can be found in Figure \ref{fig:render-case3}.

 \noindent \textbf{Scale \& Size Error} \\
The scale of axes or the size of elements is inaccurate, leading to incorrect numerical range representation or visual proportion.  Specific examples can be found in Figure \ref{fig:render-case4}.
\begin{figure}[htbp]
    \centering
    \begin{subfigure}[b]{0.23\textwidth}
        \centering
        \includegraphics[width=\textwidth]{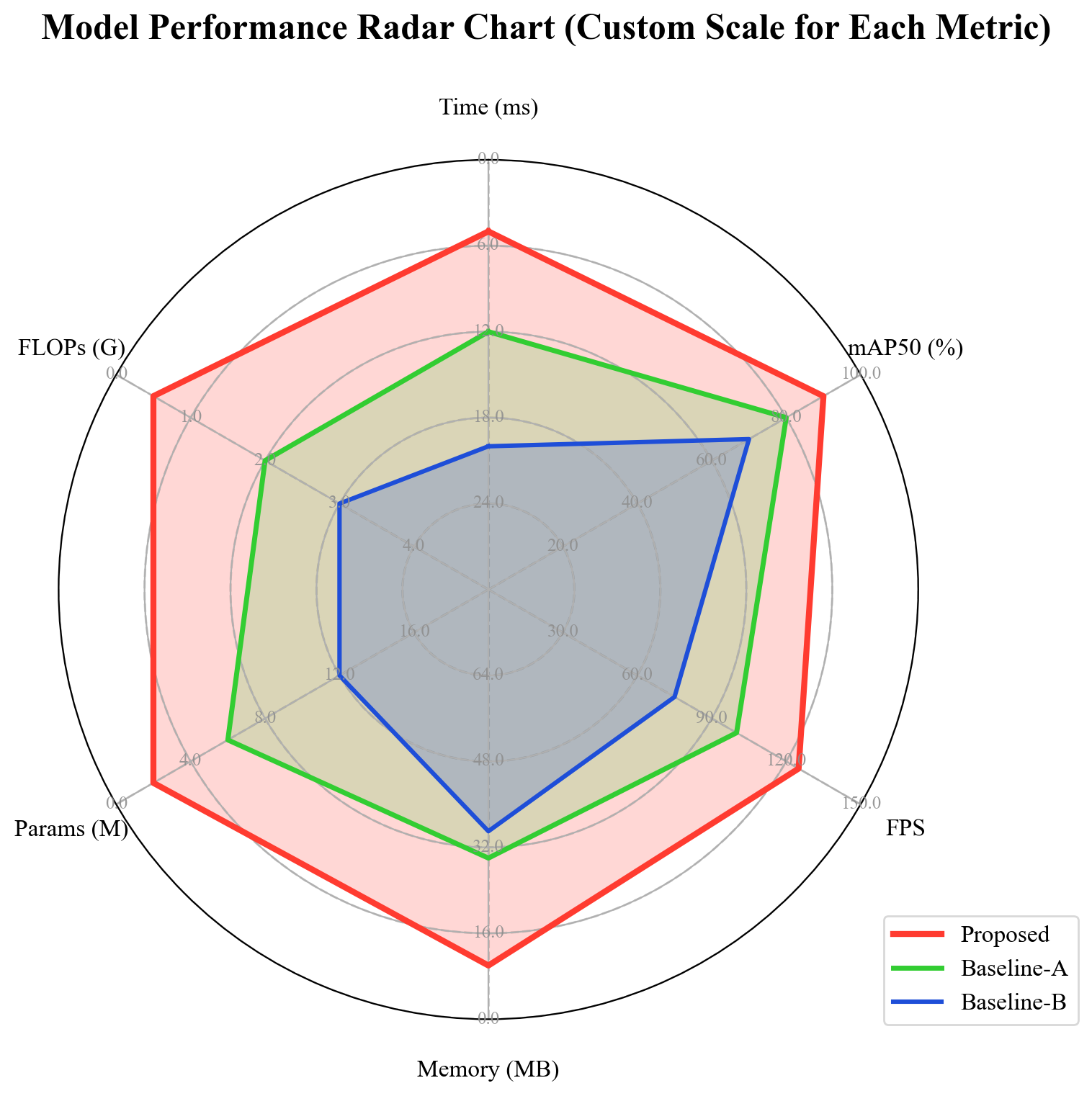}
        \caption{Ground-truth plot}
        \label{fig:code-related-errors}
    \end{subfigure}
    \hfill
    \begin{subfigure}[b]{0.23\textwidth}
        \centering
        \includegraphics[width=\textwidth]{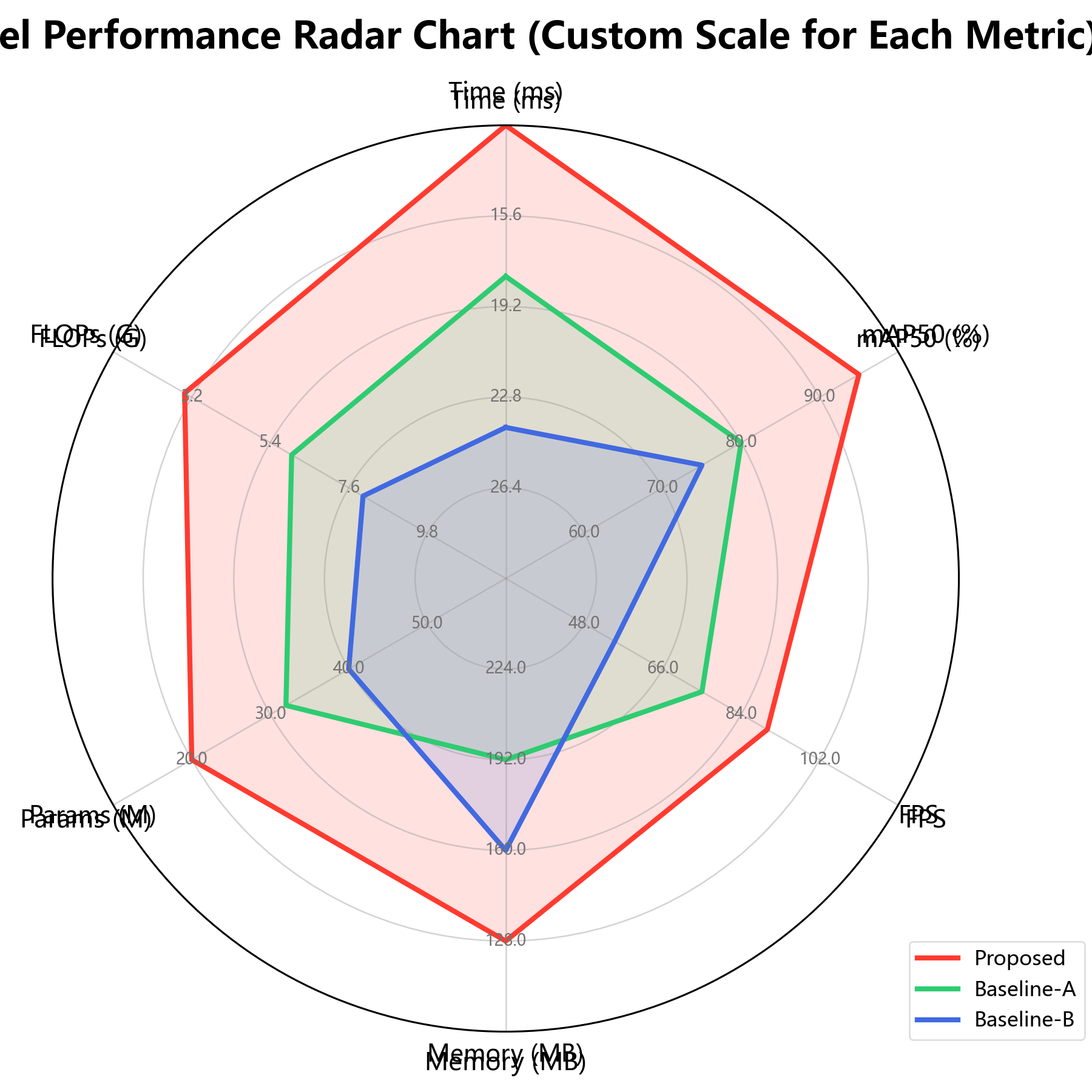}
        \caption{Generated plot}
        \label{fig:render-case4}
    \end{subfigure}
    \caption{Error Case4. In this example, the primary error lies in axis misconfiguration, which leads to incorrect data representation. In addition, element duplication and overlap are observed. }
    \label{fig:render-case4}
\end{figure}

\begin{figure}[htbp]
    \centering
    \begin{subfigure}[b]{0.23\textwidth}
        \centering
        \includegraphics[width=\textwidth]{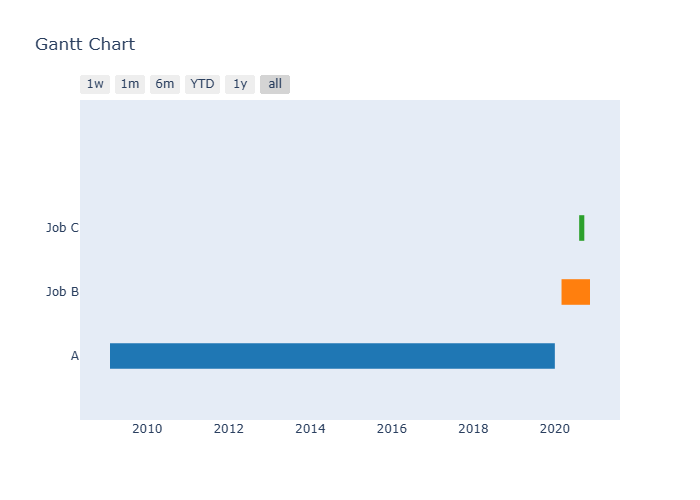}
        \caption{Ground-truth plot}
        \label{fig:code-related-errors}
    \end{subfigure}
    \hfill
    \begin{subfigure}[b]{0.23\textwidth}
        \centering
        \includegraphics[width=\textwidth]{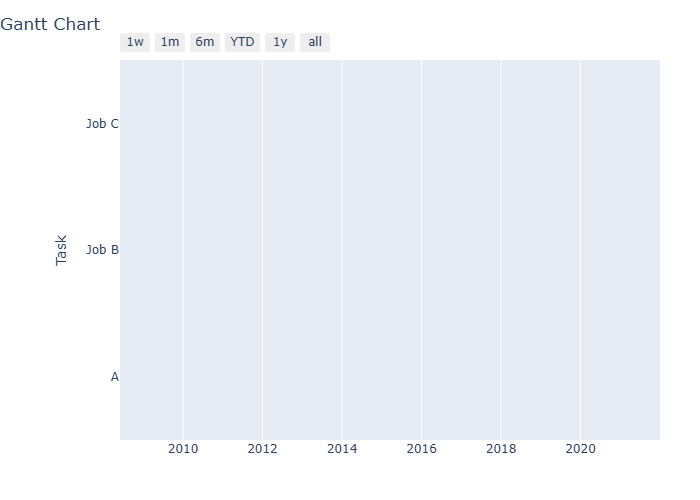}
        \caption{Generated plot}
        \label{fig:render-case4}
    \end{subfigure}
    \caption{Error Case5. In this example, the main error is that all the critical components in the figure are missing. }
    \label{fig:render-case5}
\end{figure}

 \noindent \textbf{Element Missing Error} \\
Essential visual elements such as legends, labels, or data points are not generated. Specific examples can be found in Figure \ref{fig:render-case5}.
\begin{figure}[htbp]
    \centering
    \begin{subfigure}[b]{0.23\textwidth}
        \centering
        \includegraphics[width=\textwidth]{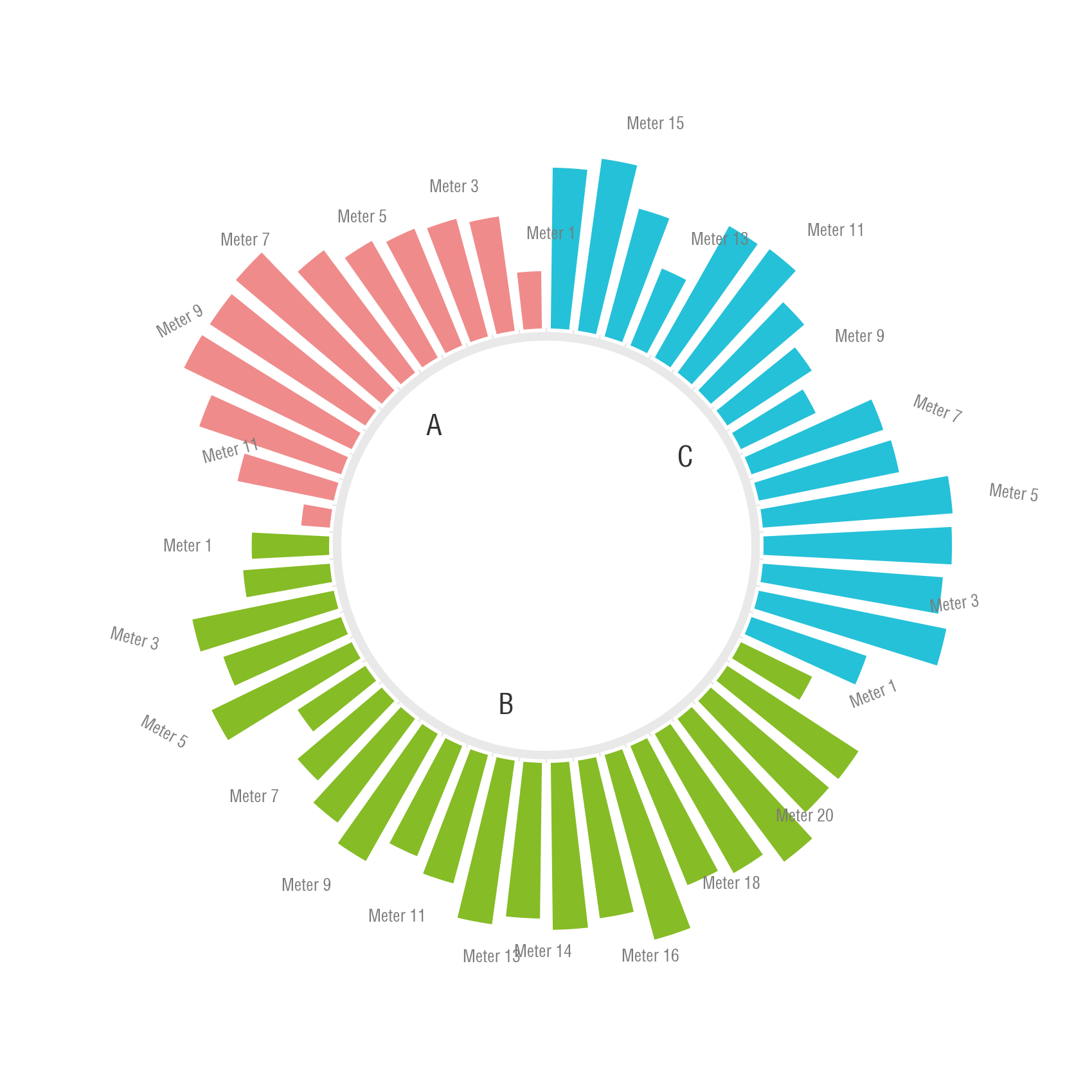}
        \caption{Ground-truth plot}
        \label{fig:code-related-errors}
    \end{subfigure}
    \hfill
    \begin{subfigure}[b]{0.23\textwidth}
        \centering
        \includegraphics[width=\textwidth]{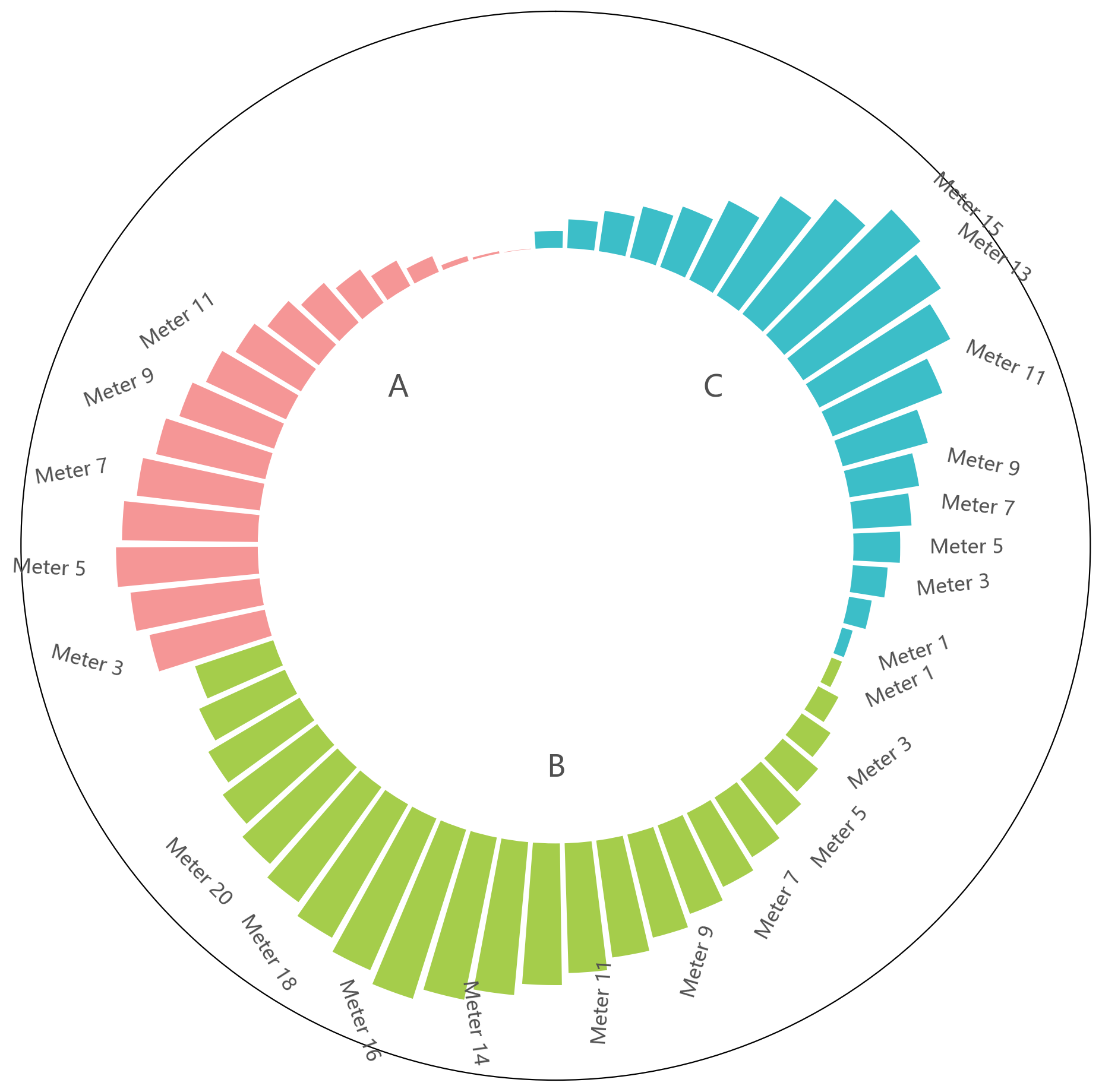}
        \caption{Generated plot}
        \label{fig:render-case4}
    \end{subfigure}
    \caption{Error Case5. In this example, the main error is that the x-axis in the chart above has extra labels that do not exist in the original plot.}
    \label{fig:render-case6}
\end{figure}

 \noindent \textbf{Element Incorrect Error} \\
Elements that do not exist in the original image are erroneously generated (e.g., extra curves or annotations). Specific examples can be found in Figure \ref{fig:render-case6}.

\section{Ethical Considerations, Societal Impact and Scientific Value}
\textbf{Ethical Considerations.}
The design of Plot-Gen Benchmark places a strong emphasis on ethical issues. Our data is primarily sourced from scientific research and publicly available online examples, and undergoes rigorous manual review to ensure that no sensitive personal information is disclosed, while minimizing potential bias or misleading content. This approach guarantees the ethical integrity of the dataset and enhances the reproducibility and credibility of the research.

\textbf{Societal Impact and Scientific Value.}
By providing a systematic, science-driven evaluation platform, Plot-Gen Benchmark contributes to advancing VLMs in scientific visualization, data analysis, and knowledge dissemination. The benchmark encourages developers to focus on cross-library execution reliability, visual consistency, and semantic interpretability of generated code, laying the foundation for responsible and verifiable intelligent visualization tools, and holding significant value for the scientific community.
\end{document}